\documentclass[11pt,twoside]{atmp}
\usepackage{amssymb}
\usepackage{amscd}
\usepackage{amsthm}
\usepackage[arrow,matrix,curve]{xy}
\usepackage{graphicx}
\newcommand\datver[1]{\def\datverp%
   {\par\boxed{\boxed{\text{Run: \today}}}}}

\newcommand\boxb[1]{\square_b}

\newcommand\pt{\text{\textup{pt}}}

\renewcommand{\theequation}{\arabic{equation}}
\newcommand\paperbody%
          {\renewcommand{\theequation}{\thesection.\arabic{equation}}}

\newtheorem{theorem}{Theorem}[section]

\newtheorem*{setup}{Basic Setup}
\newtheorem{lemma}[theorem]{Lemma}
\newtheorem{proposition}[theorem]{Proposition}
\newtheorem{corollary}[theorem]{Corollary}

\theoremstyle{remark}

\newtheorem{remark}[theorem]{Remark}

\newcommand\co{\colon\,}
\newcommand\Ind{\operatorname{Ind}}

\newcommand\Ch{\operatorname{Ch}}

\newcommand\lp{\textup{(}}
\newcommand\rp{\textup{)}}

\newcommand\Fg{\mathfrak{g}}

\newcommand\cH{\mathcal{H}}
\newcommand\cK{\mathcal{K}}

\newcommand\CC{\mathbb C}

\newcommand\PP{\mathbb P}

\newcommand\RR{\mathbb R}

\newcommand\ZZ{\mathbb Z}

\newcommand\bbC{\mathbb C}

\newcommand\bbP{\mathbb P}

\newcommand\bbR{\mathbb R}

\newcommand\bbT{\mathbb T}
\newcommand\bbZ{\mathbb Z}
\newcommand\bZ{\bbZ}
\newcommand\bR{\bbR}
\newcommand\bT{\bbT}
\newcommand\bC{\bbC}
\newcommand\bP{\bbP}

\newcommand{\bF}{{\mathbb F}_2}

\newcommand\bE{\mathbf E}
\newcommand\wE{\widetilde{\mathbf E}}
\newcommand\bh{\mathbf h}
\newcommand\bc{\mathbf c}
\newcommand\bch{\widehat{\mathbf c}}
\newcommand\wf{\widehat{f}}
\renewcommand\wp{\widehat{\varphi}}
\newcommand\pr{\text{pr}}

\newcommand{\twedge}{{\textstyle \bigwedge}}
\newcommand{\TT}{{\mathbb T}}
\newcommand\cont{\text{\textup{cont}}}
\newcommand\Lie{\text{\textup{Lie}}}

\newcommand\Ad{\operatorname{Ad}}

\newcommand\Hom{\operatorname{Hom}}
\newcommand\Aut{\operatorname{Aut}}

\newcommand{\Br}{\operatorname{Br}}
\newcommand{\Maps}{\operatorname{Maps}}
\newcommand{\wR}{\widetilde R}

\newcommand\Ca{$C^*$-algebra}
\newcommand\CTalg{continuous-trace algebra}

\renewcommand{\theequation}{\arabic{equation}}
\datver{}
\begin{document}
\title[T-duality via noncommutative topology, II]{T-duality for torus 
bundles\\with H-fluxes\\
via noncommutative topology, II:\\the high-dimensional case\\
and the T-duality group}

\author{Varghese Mathai}
\address{Department of Pure Mathematics, University of Adelaide,\\
Adelaide, SA 5005, Australia}
\email{mathai.varghese@adelaide.edu.au}
\urladdr{http://internal.maths.adelaide.edu.au/people/vmathai/}
\author{Jonathan Rosenberg}
\thanks{VM was supported by the Australian Research
Council.
JR was partially supported by NSF Grants DMS-0103647
and DMS-0504212, and thanks
the Department of Pure Mathematics of the University of 
Adelaide for its hospitality in January 2004 and August 2005, 
which made this collaboration possible.}
\address{Department of Mathematics,
University of Maryland,\\
College Park, MD 20742, USA}
\email{jmr@math.umd.edu}
\urladdr{http://www.math.umd.edu/\raisebox{-.6ex}{\symbol{"7E}}jmr}
\dedicatory{\datverp}
\begin{abstract}
We use noncommutative topology to study T-duality
for principal torus bundles with H-flux.
We characterize precisely when there is a ``classical''
T-dual, i.e., a dual bundle with dual H-flux, and
when the T-dual must be ``non-classical,'' 
that is, a continuous field of noncommutative tori. 

The duality comes with an
isomorphism of twisted $K$-theories, required for matching
of D-brane charges, just as in the classical
case. The isomorphism of twisted cohomology which one gets in
the classical case is replaced in the non-classical
case by an isomorphism of twisted cyclic homology.

An important part of the paper contains a detailed analysis of the 
classifying space for topological T-duality, as well as the T-duality 
group and its action. The issue of possible non-uniqueness
of T-duals can be studied via the action of the T-duality group.

\end{abstract}
\maketitle

T-duality is a duality of type II string theories that
involves exchanging a theory compactified on a torus with a theory
compactified on the dual torus. 
The T-dual of a  type II string theory compactified on a circle,
in the presence of a  
topologically nontrivial NS 3-form H-flux, was analyzed in special cases in
\cite{AABL, Alv12, DLP, GLMW, KSTT}. There it was observed that T-duality
changes not only the H-flux, but also the spacetime topology.  
A general formalism for dealing with T-duality for
compactifications arising from a free circle action
was developed in \cite{BEMa}.  This formalism
was shown to be compatible with two physical constraints:
(1) it respects the local Buscher rules \cite{Bus1, Bus2},
and (2) it yields an isomorphism on twisted $K$-theory, in which
the Ramond-Ramond charges and fields take their values \cite{MMSi, Wa,
Wb, MM, BM, Mkthy}.
It was shown in \cite{BEMa} that T-duality exchanges the first Chern class 
with the fiberwise integral of the H-flux, thus giving a formula for 
the T-dual spacetime topology. The purpose of this paper is
to extend these results to the case of torus bundles of
higher rank.

It is common knowledge that noncommutative tori occur naturally 
in string theory and in M-theory compactifications \cite{SW, CDS}.
This paper derives another instance where
they make a natural appearance, in the sense that if we start with 
a classical spacetime that is a principal torus bundle with H-flux, 
then the T-dual is sometimes a continuous field of noncommutative tori,
and we characterize exactly when this happens.

The first part of the paper forms a sequel to our earlier paper
\cite{MR} on T-duality for principal torus bundles with an
H-flux. In that paper, we dealt primarily with the case of fibers which
are tori of dimension $2$, in which case every principal bundle
is T-dualizable, though not necessarily in the ``classical'' sense; in fact,
if the integral of the H-flux over the torus fibers is non-trivial in 
cohomology, 
then the T-dual of such a principal torus bundle with H-flux is a 
continuous field of noncommutative tori. A similar phenomenon
was also noticed in \cite{LNR}, in fact for one of the same
examples studied in \cite{MR} (a trivial $T^2$-bundle over $S^1$,
but with non-trivial flux).
In this paper, we consider principal torus bundles of arbitrary dimension.
Still another phenomenon appears: there are some principal bundles
with H-flux which are not dualizable even in our more general sense
(where the T-dual is allowed to be a $C^*$-algebra). The 
simplest case of this phenomenon is the 3-dimensional torus, 
considered as a 3-torus bundle over a point, with H-flux chosen 
to be a non-zero integer multiple of the volume 3-form on the torus. 
More precisely, a given principal torus bundle with 
H-flux, over a base $Z$, is T-dualizable in our generalized sense if 
and only if the 
restriction of the H-flux to a torus fiber $T$ is trivial in cohomology. 
Moreover, the T-dual of such a
principal torus bundle with H-flux is ``non-classical'' if and only if 
the push-forward of the flux in $H^1(Z, H^2(T))$ is non-trivial. 
In \cite{BHMc}, the results of \cite{MR}
and of this paper were applied and extended, and it was shown that every 
principal torus bundle is T-dualizable in an even more general sense,
where the T-dual is a field of {\em non-associative} tori, i.e., taking 
us out of the category of $C^*$-algebras!

The other main part of this paper contains the analysis of the
classifying space for topological T-duality and of the 
T-duality group, as well as its action. The theory divides naturally 
into two cases, when there is a classical T-dual, and when there is
only a non-classical T-dual, and the T-duality group respects these
two cases. One somewhat unexpected result is that, when $n=2$,
the classifying space for torus bundles with H-flux splits as a 
product, one factor corresponding to the classical case and one
factor corresponding to the non-classical case.
For a rank $n$ torus bundle with $H$-flux, the T-duality
group is $GO(n, n;\bbZ)$, and acts by homotopy automorphisms of the
classifying space for classically dualizable bundles. Study of
this group enables us to understand puzzling instances of
non-uniqueness of T-duals.

Some of the results of this paper were announced in \cite{MR1}.

\section{{Notation and review of results from \cite{MR}}}
\label{sec:framework}

We begin by reviewing the precise mathematical framework from \cite{MR}.
We assume $X$ (which will be
the spacetime of a string theory) is a (second-countable)
locally compact Hausdorff space.  
In practice it will usually be a compact manifold, though 
we do not need to assume this. But we assume that $X$ is finite-dimensional 
and has the homotopy type of a finite CW-complex, in order 
to avoid some pathologies. We assume $X$ comes with a free
action of a torus $T$; thus (by the Gleason slice theorem
\cite{Gl}) the quotient map $p\co X\to Z$
is a principal $T$-bundle.

All {\Ca}s and Hilbert spaces in this paper will be over $\bbC$.
A \emph{{\CTalg} $A$ over $X$} is a particularly nice type I
{\Ca} with Hausdorff spectrum $X$ and good local
structure (the ``Fell condition'' \cite{F} --- that there
are continuously varying rank-one projections in a neighborhood
of any point in $X$). We will always assume $A$ is
separable; then a basic structure theorem of Dixmier and Douady
\cite{DD} says that after stabilization (i.e., tensoring
by $\cK$, the algebra of compact operators on an infinite-dimensional
separable Hilbert space $\cH$), $A$ becomes \emph{locally}
isomorphic to $C_0(X,\cK)$, the continuous $\cK$-valued
functions on $X$ vanishing at infinity. However, $A$ need not
be \emph{globally} isomorphic to $C_0(X,\cK)$, even after
stabilization. The reason is that a stable {\CTalg} is the
algebra of sections (vanishing at infinity) of a bundle
of algebras over $X$, with fibers all isomorphic to $\cK$.
The structure group of the bundle is $\Aut \cK\cong PU(\cH)$,
the projective unitary group $U(\cH)/\TT$. Since $U(\cH)$ is
contractible and the circle group $\TT$ acts freely on it,
$PU(\cH)$ is an Eilenberg-Mac Lane $K(\bbZ,2)$-space,
and thus bundles of this type are classified by
homotopy classes of continuous maps from $X$ to
$BPU(\cH)$, which is a $K(\bZ,3)$-space, or in other words by $H^3(X,\bZ)$. 
In this way, one can show that the {\CTalg}s over $X$, modulo
Morita equivalence over $X$, naturally form a group under the
operation of tensor product over $C_0(X)$, called the
\emph{Brauer group} $\Br(X)$, and that this group is
isomorphic to $H^3(X,\bbZ)$ via the Dixmier-Douady class.
Given an element $\delta\in H^3(X,\bbZ)$, we denote by
$CT(X,\delta)$ the associated stable {\CTalg}. (If 
$\delta=0$, this is simply $C_0(X,\cK)$.) The 
(complex topological) $K$-theory
$K_\bullet(CT(X,\delta))$ is called the \emph{twisted
$K$-theory} \cite[\S2]{RosCT} of $X$ with twist $\delta$, denoted
$K^{-\bullet}(X,\delta)$. When $\delta$ is torsion,
twisted $K$-theory had earlier been considered by
Karoubi and Donovan \cite{KD}. When $\delta=0$, twisted $K$-theory 
reduces to ordinary $K$-theory (with compact supports).

Now recall we are assuming $X$ is equipped with a free $T$-action
with quotient $X/T=Z$. (This means our theory is ``compactified
along tori'' in a way reflecting a global symmetry
group of $X$.) In general, a group action on $X$ need not
lift to an action on $CT(X,\delta)$ for any value
of $\delta$ other than $0$, and even when such a lift
exists, it is not necessarily essentially unique.
So one wants a way of keeping track of what lifts are
possible and how unique they are.
The \emph{equivariant Brauer group} defined in
\cite{CKRW} consists of equivariant Morita
equivalence classes of {\CTalg}s over $X$ equipped with
group actions lifting the action on $X$. 
Two group actions on the same stable {\CTalg} over $X$
define the same element in the equivariant Brauer group
if and only if they are outer conjugate. (This implies
in particular that the crossed products are isomorphic.
However, it is perfectly possible for the crossed products
to be isomorphic even if the actions are \emph{not} outer
conjugate.)  
Now let $G$ be the universal cover of the torus $T$,
a vector group. Then $G$ also acts on $X$ via the
quotient map $G\twoheadrightarrow T$ (whose
kernel $N$ can be identified with the free abelian
group $\pi_1(T)$). In our situation there are three Brauer
groups to consider: $\Br(X)\cong H^3(X,\bbZ)$,
$\Br_T(X)$, and $\Br_G(X)$, but
$\Br_T(X)$ is rather uninteresting, as it is naturally
isomorphic to $\Br(Z)$ \cite[\S6.2]{CKRW}. 
Again by \cite[\S6.2]{CKRW}, the natural ``forgetful map''
(forgetting the $T$-action) $\Br_T(X)\to \Br(X)$ 
can simply be identified with
$p^*\co \Br(Z)\cong H^3(Z,\bbZ) \to H^3(X,\bbZ)\cong \Br(X)$.

The basic setup from \cite{MR} is
\begin{setup}[{\cite[3.1]{MR}}]
\label{assumptions}
A spacetime $X$ compactified over a torus $T$ will
correspond to a space $X$ {\lp}locally compact, finite-dimensional,
homotopically finite{\rp} equipped with a free
$T$-action. Without essential loss of generality, we may as well
assume that $X$ is connected. The quotient map $p\co X\to Z$
is a principal $T$-bundle.  The NS 3-form
$H$ on $X$ has an integral cohomology class $\delta$ which
corresponds to an element of $\Br(X)\cong H^3(X,\,
\bbZ)$. A pair $(X,\delta)$ will be a candidate for
having a $T$-dual when the $T$-symmetry of $X$ lifts
to an action of the vector group $G$ on $CT(X,\delta)$, or in
other words, when $\delta$ lies in the image of the
forgetful map $F\co \Br_G(X)\to \Br(X)$.
\end{setup}

Recall from \cite{MR} that
if $T$ is a torus of dimension $n$, so that $G\cong \bbR^n$,
we have the following facts.  Here we 
denote by $H^\bullet_M(G,A)$ the cohomology of the topological
group $G$ with coefficients in the topological $G$-module
$A$, as defined in \cite{Mcoh}. This is sometimes
called ``Moore cohomology'' or ``cohomology with Borel cochains.''
We denote by $H^\bullet_\Lie(\Fg,A)$ the Lie algebra cohomology of
the Lie algebra $\Fg$ of $G$ with coefficients in a module $A$.
\begin{theorem}[{\cite[Theorem 4.4]{MR}}] 
\label{thm:CKRWimproved}
Suppose $G\cong \bbR^n$ is a vector group and
$X$ is a locally compact $G$-space {\lp}satisfying our
finiteness assumptions{\rp}. Then there is an exact sequence:
\[
\xymatrix@C-1.2pc{
H^2(X,\bbZ)\ar[r]^(.4){d_2''} & H^2_M(G, C(X,\TT)) \ar[r]^(.65)\xi & 
\Br_G(X) \ar[r]^F & H^3(X,\bbZ) \ar[r]^(.4){d_3}
& H^3_M(G, C(X,\TT)).}
\]
\end{theorem}
\begin{lemma}[{\cite[equation (5)]{MR}}] 
\label{lem:Moorecoh}
If $X$ is as above, then 
\[
H^\bullet_M(G, C(X,\TT)) \cong H^\bullet_M(G, C(X,\RR))
\]
for $\bullet>1$.
\end{lemma}
\begin{theorem}[{\cite[Cor,\ III.7.5]{Gu}}, 
quoted in {\cite[Corollary 4.7]{MR}}] 
\label{thm:vanEst}
If $G$ is a vector group with  Lie algebra $\Fg$,
and if $A$ is a $G$-module
which is a complete metrizable topological vector space,
then $H^\bullet_{\cont}(G,A)\cong H^\bullet_{\Lie}(\Fg, A_\infty)$.
{\lp}Here $A_\infty$ is the submodule of smooth vectors for the action
of $G$.{\rp}
In particular, it vanishes for $\bullet >\dim G$.
\end{theorem}

\section{Main mathematical results}
\label{sec:results}

Now we are ready to start on the main results. First we begin with a
lemma concerning the computation of a certain Moore cohomology
group. This lemma is quite similar to \cite[Lemma 4.9]{MR}.
\begin{lemma}
\label{lem:Liecohom}
If $G$ is a vector group and $X$ is a $G$-space {\lp}with
a full lattice subgroup $N$ acting trivially{\rp} as in the Basic
Setup above, then the maps $p^*\co C(Z,\bbR)\to C(X,\bbR)$
and ``averaging along the fibers of $p$''
$\int\co C(X,\bbR)\to  C(Z,\bbR)$ {\lp}defined
by $\int f(z) = \int_T f(g\cdot x) \, dg$, where
$dg$ is Haar measure on the torus $T$ and we choose 
$x\in p^{-1} (z)${\rp} induce isomorphisms
\[
H^\bullet_M(G,C(X,\bbR)) \leftrightarrows
H^\bullet_M(G,C(Z,\bbR)) \cong C(Z,\bbR)\otimes \twedge^\bullet(\Fg^*)
\]
which are inverses to one another.
\end{lemma}
\begin{proof}
We apply Lemma \ref{lem:Moorecoh} and Theorem \ref{thm:vanEst}. 
Note that the $G$-action
on $C(Z,\bbR)$ is trivial, so every element of $C(Z,\bbR)$
is smooth for the action of $G$.  But for any real vector space $V$
with trivial $G$-action,
\[
H^\bullet_M(G,V)\cong H^\bullet_{\Lie}(\Fg,V) \cong  
H^\bullet_{\Lie}(\Fg, \bbR)\otimes V \cong  \twedge^\bullet(\Fg^*)
\otimes V.
\]

Clearly $\int\!\circ \,p^*$ is the identity on $C(Z,\bbR)$,
so we need to show $p^*\circ \int$ induces an isomorphism
on cohomology of $C(X,\bbR)$.  The calculation turns out to be local,
so by a Mayer-Vietoris argument we can reduce to the
case where $p$ is a trivial bundle, i.e., $X=(G/N)\times Z$,
with $G$ acting only on the first factor.
The smooth vectors in $C(X,\bbR)$ for the action of
$G$ can then be identified with $C(Z,C^\infty(G/N))$.
So we obtain
\[
H^\bullet_M\bigl(G,C(X,\bbR)\bigr) 
\cong H^\bullet_{\Lie}\bigl(\Fg,C(Z,C^\infty(G/N))\bigr)
\cong C\Bigl(Z, H^\bullet_{\Lie}\bigl(\Fg, C^\infty(G/N)\bigr)\Bigr),
\]
with the cohomology moving inside since $G$ acts trivially
on $Z$.  However, we have by \cite[Ch.\ VII, \S2]{BW} that
\[
H^\bullet_{\Lie}\bigl(\Fg, C^\infty(G/N)\bigr)
\cong
H^\bullet_M(N,\bbR)\cong \twedge^\bullet (\Fg^*).
\]
\end{proof}

The first main result
generalizes the result in \cite{MR} that says that the forgetful map
$F\co \Br_G(X)\to\Br(X)$ is surjective when $\dim G\le 2$. In higher
dimensions, $F$ need not be surjective, but we characterize its image.
\begin{theorem}
\label{thm:imageofF}
Let $T$ be a torus, $G$ its universal covering, and
$p\co X\to Z$ be a principal $T$-bundle, as in the Basic Setup of section
\textup{\ref{sec:framework}}. Then the image of the forgetful map
$F\co \Br_G(X)\to H^3(X,\ZZ)$ is precisely the kernel of the
map $\iota^*\co H^3(X,\ZZ)\to H^3(T,\ZZ)$ induced by the inclusion
$\iota\co T\hookrightarrow X$ of a torus fiber into $X$.
\end{theorem} 
\begin{proof}
There are two distinct parts to the proof.
First we show that if $\delta\in H^3(X,\ZZ)$ and $\iota^*(\delta)\ne 0$, 
then $\delta$ cannot be
in the image of $F$. For this (since we can always restrict to a 
$G$-invariant set), it suffices to consider the case where $X=T$ with
$T$ acting simply transitively. Then if $\delta$ lies in the
image of $F$, that means the corresponding principal $PU$-bundle
$E\to T$ (with classifying map $\delta$) carries an action of $G$ for
which the bundle projection is $G$-equivariant.  This action corresponds
to an integrable way of choosing horizontal spaces in the tangent
bundle of $E$, or in other words, to a flat connection on $E$.  That
means, since $T=G/N$ with $N$ free abelian, that the bundle,
together with its $G$-action, comes
from a homomorphism $\rho\co \pi_1(T)=N\to PU$, which we can regard as a 
projective unitary representation of $N$.
In other words, if $A=CT(T,\delta)$ carries an action of $G$ compatible
with the transitive action of $G$ on $T$, then the action is induced
from the action of $N$ on $\cK$ associated to a projective unitary 
representation $\rho$ of $N$. So we have a surjective
``induction'' homomorphism $ H^2_M(N,\TT)\cong \Br_N(\pt)\to \Br_G(G/N)$. 

Consider the
Mackey obstruction $M(\rho)\in H^2_M(N,\TT)\cong \twedge^2(N^*)\otimes \TT$,
which is the class of the $N$-action on $\cK$ as an element of $\Br_N(\pt)$.
Since $H^2_M(N,\TT)$ is generated by product cocycles
$\omega_1\otimes \cdots \otimes \omega_k\otimes 1$, where $2k\le n=\dim G$,
$N$ splits as $N_1\times \cdots N_k\times N'$ with each $N_j$ of rank
$2$, and with $\omega_j$ skew and non-degenerate on $N_j$, it is
enough to show $\delta$ is trivial in such a case. But in this case,
if we let $G_j=N_j\otimes_\ZZ\RR$, $G'=N'\otimes_\ZZ\RR$, then
$A=\Ind_N^G(\cK, \rho)$ evidently splits as an ``external tensor product''
$\Ind_{N_1}^{G_1}(\cK, \omega_1)\otimes \cdots\otimes 
\Ind_{N_k}^{G_k}(\cK, \omega_k)\otimes \Ind_{N'}^{G'}(\cK, 1)$.
Since $\Ind_{N'}^{G'}(\cK, 1)\cong C(G'/N')\otimes\cK$ has
trivial Dixmier-Douady class,
the Dixmier-Douady class of $A$ is thus determined by the
Dixmier-Douady classes of continuous-trace algebras over the
$2$-tori $G_j/N_j$. These
of course have to be trivial, since $H^3(T^2)=0$. This completes the
first half of the proof.

For the second half of the proof, we have to show that if $\iota^*\delta=0$,
then $\delta$ is in the image of the forgetful map.
We apply Theorem \ref{thm:CKRWimproved}. This says it suffices to show that
$\delta$ is in the kernel of the map
\[
d_3\co  H^3(X,\bbZ) \to  H^3_M(G, C(X,\TT)). 
\]
from that theorem.  By
Lemma \ref{lem:Liecohom}, the calculation of the Moore
cohomology group is local (in the base $Z$), and leads to a natural isomorphism
\[
H^3_M(G, C(X,\TT)) \cong C(Z,\bbR) \otimes H^3(T,\bbR).
\]
Thus $d_3$ must be a map $H^3(X,\bbZ) \to  C(Z,\bbR) \otimes H^3(T,\bbR)$
which is locally defined (in $Z$) and natural for all principal $T$-bundles.
It must therefore factor through $\iota^*$, and so everything in the kernel
of $\iota^*$ is the image of $F$.
\end{proof}

\begin{theorem}
\label{thm:gennfacts}
In the  Basic Setup with $\dim G=n$ arbitrary, there is a commutative 
diagram of exact sequences:
\begin{equation*}
\xymatrix{
&H^0(Z,\twedge^2(\bbZ^n)) \ar[d]& 0\ar[d] &\\
H^2(X,\bbZ)\ar[r]^(.4){d_2''} & H^2_M(G, C(X,\bbT)) \ar[r]^(.6)\xi
\ar@{.>}[d]_a
& \ker F\ar[r]^\eta \ar[d] & 0 \\
& C(Z, H^2_M(N,\bbT)) \ar[d]_h & \Br_G(X) \ar[l]_(.35)M \ar[d]^F & \\
 & H^1(Z,\twedge^2(\bbZ^n)) \ar[d] & H^3(X,\bbZ) \ar[d]_{\iota^*}
\ar@{.>}[l]_(.35){p_!} & \\
& 0& H^3(T,\bbZ)&
}
\end{equation*}
with $M$ the Mackey obstruction map defined in \cite{PRW} and with
\begin{multline*}
h\co C(Z, H^2_M(N,\bbT)) \cong C(Z, \twedge^2(\bbZ^n)\otimes \TT)
\to H^1(Z,\twedge^2(\bbZ^n))\cong  H^1(Z,\bbZ^k), \\
k = \binom n 2,
\end{multline*}
the map sending a continuous
function $Z\to \twedge^2(\bbZ^n)\otimes \TT$ to its homotopy class. 
Here $p_!$ is defined on $\ker \iota^*\co H^3(X,\bbZ)\to H^3(T,\bbZ)$
{\lp}the dotted arrow indicates that it is not always defined on all
of $H^3(X,\bbZ)${\rp}, and there is given by  the map
to the subquotient $E_\infty^{1,2}$ of the Serre spectral sequence,
or {\lp}more informally{\rp} by
\[
H^3(X, \bbZ) \ni H \mapsto 
\Bigl(\int_{\TT_1^2}\!\! H , \, \ldots,  \int_{\TT_k^2}\!\! H\Bigr)
\in H^1(Z,\bbZ^k),
\]
where $\TT_j^2$, $j=1, \ldots, k$ run through a basis for
the possible 2-dimensional  subtori in the fibers.
\end{theorem}
\begin{proof}
Exactness of the top horizontal sequence comes from Theorem
\ref{thm:CKRWimproved}, 
and exactness of the right-hand vertical sequence comes from
Theorem \ref{thm:imageofF}.  Exactness of the left-hand vertical sequence
comes from the isomorphism $H^2_M(G, C(X,\bbT)) \cong
H^2_M(G, C(X,\bbR))$ of Lemma
\ref{lem:Moorecoh}, the calculation of $H^2_M(G, C(X,\bbR))$                    
in Lemma \ref{lem:Liecohom}, and the exact sequence
\[
0\to H^0(Z,K) \to C(Z, K\otimes\bbR) \to  C(Z, K\otimes\bbT)
\to H^1(Z, K)\to 0
\]
for $K$ a free abelian group. (In this case, $K=\twedge^2(\bbZ^n)$.)
So it remains to verify commutativity of the
two squares.  The proof is very similar to that of
\cite[Theorem 4.10]{MR}.  Commutativity of the upper square amounts to
showing that $M\circ\xi$ is the same as the map
\[
a\co H^2_M(G, C(X,\bbT)) \xrightarrow{\cong}
C(Z, \twedge^2(N)\otimes \bbR) \xrightarrow{\exp}
C(Z, \twedge^2(N)\otimes \bbT).
\]
This is immediate from the definition of $\xi$ in \cite[Theorem
5.1]{CKRW}. The harder part of the proof is commutativity of the lower
square, i.e., showing that if
$\alpha$ of $G$ on $CT(X,\delta)$, representing an element of
$\Br_G(X)$, then $h\circ M (\alpha) = p_!(\delta)$. As in
\cite[Theorem 4.10]{MR}, $h\circ M (\alpha)$ only depends 
$\delta$, not on the specific choice of $\alpha$, since
any two actions (of the sort we are considering)
on $CT(X,\delta)$ differ by an element of
$\ker F$, which by the rest of the diagram is in the image of
$H^2_M(G, C(X,\bbT))$, and thus only changes $M(\alpha)$ within its
homotopy class.

Next we show that the map $H^3(X,\bbZ)\to H^1(Z,\twedge^2(\bbZ^n))$
induced by $h\circ M$ vanishes on the subquotients
$E_\infty^{3,0}$ and $E_\infty^{2,1}$ of $H^3(X,\bbZ)$ for
the Serre spectral sequence of the fibration $T\to X \stackrel{p}{\to}
Z$. First of all, $E_\infty^{3,0}=p^* \bigl( H^3(Z,\bbZ) \bigr)$.
If $\delta=p^*(\eta)$ with $\eta\in H^3(Z,\bbZ)$, then by
\cite[\S6.2]{CKRW}, there is a $T$-action on $CT(X,\delta)$
corresponding to the $T$-action on $X$. This lifts to a $G$-action
with the discrete subgroup $N$ acting trivially, so the Mackey
obstruction class for $N$ is certainly trivial. Secondly,
$E_\infty^{2,1}$ consists (modulo $E_\infty^{3,0}$) of
classes pulled back from some intermediate space
$Y$, where $X\xrightarrow{p_1} 
Y\xrightarrow{p_2} Z$ is some 
factorization of the $T$-bundle $p\co X\to Z$ as
a composite of two principal torus bundles, with $p_1$ having
$(n-1)$-dimensional fibers $T'$ and with $p_2$ having
one-dimensional, i.e., $S^1$, fibers. But given
such a factorization and a class $\delta_Y\in Y$, there
is an essentially unique action of $\bbR$ on 
$CT(Y,\delta_Y)$ compatible with the $S^1$-action on $Y$
with quotient $Z$, because of the results of \cite[\S 4.1]{MR}.  
Pulling back from $Y$ to $X$, we get
an action of $\bbR\times T'$ on $CT(X,p_1^*\delta_Y)$,
or in other words an action of $G$ factoring through
$\bbR\times T'$. Such an action necessarily has trivial
Mackey obstruction.

Since we are assuming that $\delta$ lies in the kernel of $\iota^*\co
H^3(X,\bbZ) \to H^3(T,\bbZ)$ (otherwise, by Theorem
\ref{thm:imageofF}, there is nothing to prove), the map induced by
$h\circ M$  factors through the remaining subquotient of $H^3(Z,\bbZ)$,
viz., $E_\infty^{1,2}$. That says exactly that the map 
factors through $p_!$. We can now compute it as we did in the proof of 
\cite[Theorem 4.10]{MR} by calculating in specific cases and using
naturality. Thus the proof is concluded using Proposition
\ref{prop:Mobstrfortrivbundle} in Section \ref{sec:examples} below.
\end{proof}

\section{Applications to T-duality}
\label{sec:mainthm}

Now we are ready to apply Theorem \ref{thm:gennfacts} to
T-duality in type II string theory. 

The following is the second main result of this paper.

\begin{theorem}\label{thm:main}
Let $p\co X\to Z$ be a principal $T$-bundle
as in the Basic Setup, where $n=\dim T$ is arbitrary. Let
$\delta \in H^3(X, \bbZ)$ be an ``H-flux'' on $X$
that is the kernel of $\iota^*\co H^3(X, \bbZ)\to H^3(T, \bbZ)$,
where $\iota$ is the inclusion of a fiber. Let $k=\binom n 2$. Then:
\begin{enumerate}
\item If $p_! \delta = 0 \in H^1(Z, \ZZ^k)$, then there 
is a classical T-dual to
$(p,\delta)$, consisting of
$p^\#\co X^\# \to Z$, which is another principal $\TT^n$-bundle
over $Z$, and $\delta^\# \in H^3(X^\#, \bbZ)$,
the ``T-dual H-flux'' on  $X^\#$. One obtains
a picture of the form
\begin{equation}
\label{eq:fibdiag}
\xymatrix{
 & X\times_Z X^\# \ar[ld]_{p^*(p^\#)} \ar[rd]^{(p^\#)^*(p)} & \\
X \ar[rd]_p& & X^\# \ar[ld]^{p^\#}\\
 & Z & .}
\end{equation}
There is a natural isomorphism of twisted $K$-theory
\[
K^\bullet(X,\delta) \cong K^{\bullet+n}(X^\#,\delta^\#) .
\]
{\lp}Because of Bott periodicity, the dimension shift can be
ignored if and only if $\dim T$ is even. The shift in
odd dimensions can be understood physically in terms of a
duality between type IIA and type IIB string theories.{\rp}
\item If $p_! \delta \ne 0 \in H^1(Z, \ZZ^k)$, then 
a classical T-dual as above does \emph{not} exist.
However, there is a ``non-classical''
T-dual bundle of noncommutative tori over $Z$. It is not
unique, but the non-uniqueness does not affect its
$K$-theory, which is isomorphic to $K^\bullet(X,\delta)$
with a dimension shift of $\dim G$ mod $2$.
\end{enumerate}  
\end{theorem}
\begin{proof}
By Theorem \ref{thm:imageofF}, the assumption that
$\iota^*\delta=0$ is equivalent to saying that $\delta$ lies in the
image of $F$.

First consider the case when $p_! \delta = 0 \in H^1(Z, \ZZ^k)$.
By commutativity of the lower square in Theorem
\ref{thm:gennfacts}, we can lift $\delta\in H^3(X,\bbZ)$ to
an element $[CT(X,\delta), \alpha]$ of $\Br_G(X)$ with
$M(\alpha)$ homotopically trivial.  Then by using
commutativity of the upper square in Theorem \ref{thm:gennfacts},
we can perturb $\alpha$, without changing $\delta$, so
that $M(\alpha)$ actually vanishes. Once this is
done, the element we get in $\Br_G(X)$ is actually unique
modulo the discrete group $H^0(Z, \bigwedge^2(\ZZ^n))/d_2''(H^2(X,\ZZ))$.
On the one hand, this can be seen from \cite[Lemma 1.3]{PRW} and
\cite[Corollary 5.18]{PRW}. Alternatively, it can be
read off from Theorem \ref{thm:gennfacts}, since any
two classes in $\ker M$ mapping to the same $\delta\in 
H^3(X,\bbZ)$ differ by the image under $\xi$ of something
in $\ker a$. The non-uniqueness of the element of
$\Br_G(X)$ will be addressed in the following Proposition.
Finally, if $[CT(X,\delta), \alpha]$ has trivial Mackey
obstruction, then as explained in \cite[\S1]{PRW} and \cite[\S3]{ER},
$CT(X,\delta)\rtimes_\alpha G$ has continuous trace and
has spectrum which is another principal torus bundle
over $Z$ (for the dual torus, $\widehat G$ divided
by the dual lattice). The ``Thom isomorphism'' of
Connes \cite{CThom} gives an isomorphism between the
$K$-theory of $CT(X,\delta)$ and that of $CT(X,\delta)\rtimes_\alpha
G$, with a dimension shift of $\dim G$.  Thus we
obtain the desired isomorphism in twisted $K$-theory.

Now consider the case when
\begin{equation}\label{h-1}
p_! \delta \ne 0 \in H^1(Z, \ZZ^k).
\end{equation}
It is still true as before that we can find an element
$[CT(X,\delta), \alpha]$ in $\Br_G(X)$ corresponding to
$\delta$. But there is no classical T-dual in this situation since
the Mackey obstruction \emph{can't} be trivial, because of
Theorem \ref{thm:gennfacts}.  In fact, since any representative
$f\co Z\to \TT$ of a non-zero class in $ H^1(Z, \ZZ)$
must take on all values in $\TT$, there are necessarily
points $z\in Z$ for which the Mackey obstruction in
$H^2(\bbZ^n,\TT)\cong \TT^k$ is irrational, and hence the
crossed product $CT(X,\delta)\rtimes_\alpha G$ cannot
be type I. Nevertheless, we can view this crossed
product as a \emph{non-classical} T-dual to $(p,\delta)$.
The crossed product can be viewed as the algebra of
sections of a bundle of algebras (not locally trivial) over
$Z$, in the sense of \cite{DH}. The fiber of this
bundle over $z\in Z$ will be $C(p^{-1}(z), 
\cK(\cH)) \rtimes G \cong C(G/\ZZ^n, \cK(\cH)) 
\rtimes G \cong A_{f(z)} \otimes \cK(\cH)$,
which is Morita equivalent to the
twisted group {\Ca} $A_{f(z)} $ of the stabilizer group $\bbZ^n$ for
the Mackey obstruction class $f(z)$ at that point. In other words,
the  T-dual will be realized by a bundle of (stabilized)
\emph{noncommutative tori} fibered over $Z$. (See Figure 1.)

\begin{figure}[ht]
\includegraphics[height=1.5in]{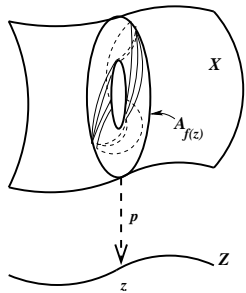}\\
\caption{In the diagram, the fiber over $z\in Z$ 
is the noncommutative torus $A_{f(z)}$,
which is represented by a foliated torus, with foliation 
angle equal to $f(z)$.}
\label{fig:folbundle}
\end{figure}

The bundle is not unique since there is no \emph{canonical}
representative $f$ for a given non-zero class in $H^1(X,\bbZ)$.
However, any two choices are homotopic, and the resulting
bundles will be in some sense homotopic to one another.
The isomorphism of the $K$-theory of the crossed product
with the twisted $K$-theory of $(X,\delta)$ again follows from
the ``Thom isomorphism'' of Connes \cite{CThom} 
\end{proof}
\begin{remark}
\label{rem:cyclichom}
An almost identical argument, with Connes' Theorem
replaced by its analogue in cyclic homology
\cite{ENN}, can be used to get an isomorphism of
twisted cohomology groups in the classical case,
and of twisted cohomology with cyclic homology
of the (smooth) dual, in the non-classical case.
\end{remark}

The next proposition corrects one small problem in Theorem
\ref{thm:main}, which also appeared in \cite[Theorem 4.13]{MR}.
Namely, it would appear that the ``classical T-dual'' in Theorem
\ref{thm:main} is not unique, since when it exists, the
lifting of $\delta$ to a element of $\Br_G(X)$ with vanishing
Mackey obstruction is not necessarily unique. We thank Bunke, Rumpf,
and Schick \cite{BS2} for noticing this. However, it turns out that
the non-uniqueness doesn't matter after all.

\begin{proposition}
\label{prop:nonuniqueness}
Given a principal $T$-bundle $p\co X\to Z$ and an
element $\delta\in H^3(X,\ZZ)$ such that $\iota^*\delta=0$
in $H^3(T,\ZZ)$ and such that $p_!(\delta)=0$ in $H^1(Z, H^2(T,\ZZ))$,
the non-uniqueness of the lift  of $\delta$ to an
element $\alpha\in \Br_G(X)$ with $M(\alpha)=0$
does not affect the classical T-dual
$p^\#\co X^\#\to Z$ or the T-dual H-flux $\delta^\#\in 
 X^\#$.
\end{proposition}
\begin{proof}
Let $A=CT(X,\delta)$. This is a continuous-trace algebra with a 
free action of the torus $T$ of rank $n$ on its spectrum. 
Let $\alpha$ be a lift of the $T$-action to an action of the universal 
cover $G$ of $T$ on $A$, such that the Mackey obstruction $M(\alpha)$
is identically $0$. Chasing the diagram in Theorem \ref{thm:gennfacts}
shows that any other such lift is of the form $\alpha\otimes\omega$,
where $\omega$ is an action of $G$ on $\cK$ defined by a
projective unitary representation of $G$ with Mackey obstruction
lying in $\ker\left(  \RR^k\cong H^2_M(G,\TT) \to
H^2_M(N,\TT)\cong \TT^k\right)$, $k=\binom n 2$.  Here again
$N$ is the lattice subgroup $\ker (G\twoheadrightarrow T)$. (Note that
tensoring with $\omega$ doesn't change the condition of vanishing of
the Mackey obstruction, because of the fact $\omega|_N$ has trivial
obstruction.)  We want to prove that the following crossed products are
isomorphic:
\begin{equation}\label{eq:induction}
A \rtimes_\alpha G \cong A \rtimes_{\alpha\otimes \omega} G. 
\end{equation}
We do this by induction on $n$. To start the induction, the result
is obvious when $n=1$, since then $\alpha$ is unique up to
exterior equivalence (see \cite[Corollary 6.1]{CKRW} and 
\cite[\S4.1]{MR}). So assume $n>1$ and suppose the result is
true for smaller values of $n$. Write $N$ as $N_1\times\ZZ$,
$G$ as $H\times\bbR$, where $H = N_1\otimes_{\ZZ}\RR
\cong \bbR^{n-1}$. Note that $H/N_1\cong \TT^{n-1}$ also acts
freely on $X$ (even though the quotient space is different from
the original $Z$).  So by inductive hypothesis, 
with $G$ replaced by $H$ in \eqref{eq:induction},
$A \rtimes_{\alpha |_H}
H \cong A \rtimes_{\alpha\otimes\omega |_H} H$. So their spectra
and Dixmier-Douady classes are the same.  In fact, since
the action of the other factor of $\bbR$ on the spectra of these
crossed products still comes from the original action of $T$ on
$\widehat A$, these spectra are homeomorphic not just as
spaces, but equivariantly as $\TT$-spaces, 
for the free action of the remaining
circle in $T$. Thus, by the case $n=1$ applied to this situation,
\[
\begin{aligned}
A \rtimes_{\alpha\otimes \omega} G 
&= (A \rtimes_{\alpha\otimes\omega |_H} H)
\rtimes_{\alpha\otimes\omega  |_\bbR} \bbR\\
&\cong (A \rtimes_{\alpha  |_H} H)
\rtimes_{\alpha|_\bbR} \bbR\\
&\cong A \rtimes_\alpha G.
\end{aligned}
\]
That completes the inductive step.
\end{proof}

\section{Examples}
\label{sec:examples}

We begin with the following explicit 
construction, where a field of noncommutative 
tori appears in the explicit T-dual.

\begin{proposition}
\label{prop:Mobstrfortrivbundle}
Let $p\co X=Z\times T\to Z$ be the trivial $T$-bundle,
where $T = \TT^n$.
Let $\beta\in H^1(Z,\bbZ^k) = [Z, \TT^k]$ be a non-zero element,
where $k=\binom n 2$, 
and let $\delta=\langle\beta, \gamma\rangle
\in H^3(X,\bbZ)$, where $\gamma \in H^2(T,  {\ZZ^k})$ 
and the brackets denote the duality pairing. 
Then there is an explicit construction of  $CT(X,\delta)$
together with an action
$\alpha$ of $G=\bbR^n$ on $CT(X,\delta)$, compatible with the
free $T$-action on $X$, for which $h\circ M(\alpha)
=\beta$. As a consequence, we get an explicit description 
of the $T$-dual to the trivial principal torus bundle with 
H-flux equal to $\delta$, that is, an explicit description of 
$CT(X,\delta) \rtimes_\alpha G$.
\end{proposition}

\begin{proof} First of all, notice that $\delta$ is in the 
the kernel of $\iota^*\co H^3(X, \bbZ)\to H^3(T, \bbZ)$,
where $\iota$ is the inclusion of a fiber.
Choose a function $f\co Z\to \TT^k$ such that $h(f)=\beta$.
For each $f(z) \in \TT^k \cong H^2(\ZZ^n, \TT)$, choose a multiplier 
$\sigma_{f(z)}$ on $\ZZ^n$ such that 
$[\sigma_{f(z)}] = f(z)\in H^2(\ZZ^n, \TT)$.
Let $\cH$ denote the Hilbert space $\ell^2(\ZZ^n)$. Then there is a 
natural projective unitary representation $\rho_{f(z)}
: \ZZ^n \to PU(\cH)$ which is the right $\sigma_{f(z)}$-regular
representation, defined by  
$\rho_{f(z)} (s) f(s') = f(s + s') 
\sigma_{f(z)}(s, s')$,
and whose Mackey obstruction is equal to $f(z)  \in H^2(\ZZ^n, \TT)$. 
Let $\ZZ^n$ act on $C(Z, \cK(\cH))$ by $\alpha$, which is given at 
the  point $z\in Z$ by $\Ad(\rho_{f(z)})$, which is 
a spectrum-fixing
automorphism of $\ZZ^n$ on $C(Z, \cK(\cH))$. 
Define the $C^*$-algebra
$$
  \begin{array}{lcl}
  B &=& {\Ind}_{\ZZ^n}^{\RR^n}\left(C(Z, \cK(\cH)), \alpha\right) 
\\[+7pt]
  &=& \left\{ \lambda: \RR^n \to C(Z, \cK(\cH)):
\lambda(t+g) = \alpha(g)( \lambda(t)),\;\; t\in \RR^n, g\in\ZZ^n \right\}.
  \end{array}
  $$
Since $\alpha$ acts trivially on the spectrum $Z$ of the inducing
algebra and $B$ is an algebra of sections of a locally trivial
bundle of {\Ca}s with fibers isomorphic to $\cK$, it follows that
$B$ is a {\CTalg} having spectrum $Z\times T$.  There is
a natural action $\xi$ of $G$ on $B$ by translation, and by 
construction, the Mackey obstruction $M(\xi)=f$.  
The computation of the Dixmier-Douady invariant of $B$
is done exactly as \cite[Proposition 4.11]{MR}, 
and is equal to $\delta=\langle\beta, \gamma\rangle
\in H^3(X,\bbZ)$ as claimed. That is, $B = CT(X,\delta)$, and so 
the T-dual is $B\rtimes_\alpha G$.
\end{proof}

The following is a special case of the above, where we can be 
even more explicit.

\begin{proposition} 
Let $Z= \bbT^k$, $T= \bbT^n$ and consider the 
trivial torus bundle $X = Z \times T \to Z$.
Let $\beta \in H^1(Z, \ZZ^k)= [\bbT^k,  \bbT^k]$ be an element of degree one,
and  $ \tau \in H^2(T,  {\ZZ^k}) \cong 
H^2(\ZZ^n, \ZZ^k)$ determine the 
central extension,
$$
1\to \ZZ^k \to H_\tau \to \ZZ^n \to 1,
$$
where $H_\tau$ is a
2-step nilpotent group over $\ZZ$. {\lp}In fact,
by \textup{\cite{PS}}, every torus bundle over a torus
is a nilmanifold with fundamental group of the form
$H_\tau$.{\rp} Then the T-dual of the trivial torus bundle 
$X= Z \times T \to Z$, with H-flux given by 
$\delta = \langle\beta, \tau\rangle \in H^3(X, \bbZ)$ 
{\lp}the brackets denote the duality pairing{\rp}, 
is the stabilized group $C^*$-algebra $C^*(H_\tau)\otimes \cK$, which
can be realized as  
a continuous field of stabilized noncommutative tori of dimension $n$
that are parameterized by $\TT^k$. 
\end{proposition}

\begin{proof}
Here we choose the identity map $I : \TT^k \to \TT^k$ as representing
the degree one element $\beta  \in H^1(Z, \ZZ^k)$.
For each  $\theta \in \TT^k \cong H^2(\ZZ^n, \TT)$,  choose a
multiplier $\sigma_\theta$ on  
$\ZZ^n$ such that $[\sigma_\theta] = \theta \in H^2(\ZZ^n, \TT)$. Let
$\cH$ denote the  
Hilbert space $\ell^2(\ZZ^n)$. Then there is a natural projective
unitary representation 
$\rho_\theta : \ZZ^n \to PU(\cH)$ which is the right
$\sigma_\theta$-regular representation,  
defined by $\rho_\theta (\gamma) f(\gamma') = f(\gamma + \gamma')
\sigma_\theta(\gamma, \gamma')$, 
and whose Mackey obstruction is equal to $\theta  \in H^2(\ZZ^n, \TT)$. 
Let $\ZZ^n$ act on $C(\TT^k, \cK(\cH))$ by $\alpha$, which is given at 
the  point $\theta$ by $\rho_\theta$. Define the $C^*$-algebra
$$
  \begin{array}{lcl}
  B &=& {\Ind}_{\ZZ^n}^{\RR^n}\left(C(\TT^k, \cK(\cH)), \alpha\right) 
\\[+7pt]
  &=& \left\{ f: \RR^n \to C(\TT^k, \cK(\cH)):
  f(t+g) = \alpha(g)( f(t)),\;\; t\in \RR^n, g\in\ZZ^n \right\}.
  \end{array}
  $$
That is, $B$ (with an implied action of $\RR^n$)
is the result of inducing a $\ZZ^n$-action on $C(\TT^k, \cK(\cH))$
from $\ZZ^n$ up to $\RR^n$.
Then $B$ is a continuous-trace {\Ca} having spectrum $\TT^{n+k}$,
and an action of $\RR^n$ whose induced action on the spectrum of $B$
is the trivial bundle $\TT^{n+k}\to \TT^k$.  
The computation of
the Dixmier-Douady invariant of $B$
is done exactly as in Proposition 4.11, \cite{MR}, 
and is equal to $\delta=\langle\beta, \tau\rangle
\in H^3(X,\bbZ)$.
The crossed product algebra
$B\rtimes \RR^n \cong C(\TT^k, \cK(\cH))\rtimes \ZZ^n$ has fiber
over $\theta \in \TT^k$ given by
$\cK(\cH) \rtimes_{\rho_\theta} \ZZ^n \cong 
A_\theta(n) \otimes \cK(\cH),$
where $A_\theta(n)$ is the noncommutative $n$-torus.
In fact, the crossed product  $B\rtimes \RR^n$
is Morita equivalent to $C(\TT^k, \cK(\cH))\rtimes\bbZ^n$
and is even isomorphic to the stabilization of this
algebra (by \cite{Green}).  Thus $B\rtimes \RR^n$ is isomorphic
to $C^*(H_\tau) \otimes \cK$, where $H_{\tau}$ is the
2-step nilpotent group over $\ZZ$ determined by $\tau \in H^2(\ZZ^n, \ZZ^k)$.
\end{proof}

\section{The classifying space for topological T-duality,\\
and action of the T-duality group}
\label{sec:classspace}

In this section, we discuss how our results are related to a
construction of Bunke and Schick \cite{BS} concerning a ``classifying space''
for topological T-duality. We will also discuss the action of the
\emph{T-duality group}.

\subsection{Review of the Bunke-Schick construction for $n=1$}
\label{sec:BS}
In the paper \cite{BS}, Bunke and Schick have pointed out
that there is a \emph{classifying space} $R$ for principal
$\TT$-bundles with $H$-flux. In other words, the set of all possible 
principal $\TT$-bundles $p\co X\to Z$ (for fixed $Z$), each equipped
with a cohomology class $\delta\in H^3(X,\ZZ)$, modulo isomorphism,
can be identified with the set of homotopy classes of maps
$Z\to R$. Bunke and Schick call $(p\co X\to Z, \delta)$ a
\emph{pair}.  They show that there is a \emph{universal pair}
$(\bE, \bh)$ over $R$, such that any pair $(p\co X\to Z, \delta)$
is pulled back from $(\bE, \bh)$ via a map $Z\to R$ (whose homotopy
class is uniquely determined).

The Bunke-Schick classifying space $R$ has an interesting
structure. It is a two-stage Postnikov system; i.e., it has exactly
two non-vanishing homotopy groups, and can be constructed as a
principal fibration
\begin{equation}
\label{eq:Postnikov}
K(\ZZ, 3) \to R \xrightarrow{q} K(\ZZ, 2) \times K(\ZZ, 2).
\end{equation}
Here $K(\ZZ, 3)$ is the classifying space for $H^3$, a space with
exactly one non-zero homotopy group and with $\pi_3=\ZZ$, and
$K(\ZZ, 2)$ is the classifying space for $H^2$, a space with
exactly one non-zero homotopy group and with $\pi_2=\ZZ$. A useful
explicit model for $K(\ZZ, 2)$ is the infinite projective space
$\CC\PP^\infty$. The homotopy type of the fibration
\eqref{eq:Postnikov} is determined by the $k$-invariant in $H^4(K(\ZZ,
2) \times K(\ZZ, 2), \ZZ)$, which in this case is the cup-product 
$\bc\cup\bch$, where $\bc$ and $\bch$ are the canonical generators in
$H^2$ of the two $K(\ZZ, 2)$ factors. The two classes $\bc$ and $\bch$
play different, but symmetrical, roles. Let $\pr_1$ and $\pr_2$ be the
two projections $K(\ZZ, 2) \times K(\ZZ, 2)\to K(\ZZ, 2)$. Then
given a map $\varphi\co Z\to R$, with $\varphi^*(\bE, \bh) =
(p\co X\to Z, \delta)$, $\pr_1\circ q \circ \varphi\co  Z\to
K(\ZZ, 2)$ classifies the Chern class of the $\TT$-bundle $p\co X\to
Z$, while $\pr_2\circ q \circ \varphi\co  Z\to
K(\ZZ, 2)$ classifies $p_!(\delta)$.

Now T-duality replaces a pair $(p\co X\to Z, \delta)$ with a dual
pair $(p^\#\co X^\#\to Z, \delta^\#)$, where the Chern class of
$p^\#\co X^\#\to Z$ is $p_!(\delta)$, and $(p^\#)_!(\delta^\#)$ is the
Chern class of the original bundle $p\co X\to Z$. Thus we can
understand T-duality in terms of a self-map 
$\#\co R \circlearrowleft
$ which comes from interchanging the two copies of $K(\ZZ, 2)$ in the
fibration \eqref{eq:Postnikov}. (Note that since
$\bc\cup\bch=\bch\cup\bc$, this preserves the $k$-invariant and thus
extends to a self-map of $R$, with $(\#)^2$ homotopic to the identity.)

Using this point of view, we can now explain the notion of the
\emph{T-duality group} (still in the relatively trivial case of $n=1$).
Good
references include \cite{GPR} (for the state of the theory up to 1994)
and \cite{Hull} for something more current. The physics literature 
mostly talks in terms of an $O(n,n;\ZZ)$ symmetry, which when
$n=1$ degenerates to $O(1,1;\ZZ)=\{\pm 1\}\times \{\pm 1\}$. If 
this group is to act by T-duality symmetries
for general circle bundles, then it should operate on
$R$ in a way that induces this symmetry, and indeed it does.

The action of $GL(n,\ZZ)$ as discussed in \cite{BS} amounts to the
following. Given a principal $n$-torus bundle $p\co X\to Z$, we can
make a new principal $n$-torus bundle out of the same underlying
spaces $X$ and $Z$ by twisting the free action of $T=\TT^n$ on $X$ by
an element $g\in \Aut(\TT^n)=GL(n,\ZZ)$. In other words, for $t\in T$,
we define $x\cdot_g t$ to be $x\cdot g(t)$. (Here $\cdot$ is the
original free right action of $T$ on $X$, with quotient space $Z$, and
$\cdot_g$ is the new twisted action.) In the case where $n=1$ and
$g=-1$ is the non-trivial element of $GL(1,\ZZ)$, this twisting
changes the sign of the Chern class of the bundle $p$. It also changes
the sign of $p_!(\delta)$, since $\delta$ does not change but the
orientation of the fibers of $p$ is reversed (and thus the definition
of integration along the fibers changes by a sign). In view of what we
said about $\bc$ and $\bch$, that means that the action of the
twisting on $R$ reverses the sign of both $\bc$ and $\bch$. Since
$(-\bc)\cup(-\bch)= \bc\cup\bch$, once again the $k$-invariant of
\eqref{eq:Postnikov} is preserved and we get a well defined action of
$GL(1,\ZZ)$ on $R$. This action can be explicitly realized using the
complex conjugation map on each copy of $\CC\PP^\infty = K(\ZZ, 2)$.

Bunke and Schick do not discuss (in this paper---they do go into it
in the sequel paper, \cite{BS2}) the action of $O(n,n;\ZZ)$, but in the
case $n=1$ we can explain this as follows. First of all, $GL(n,\ZZ)$
is to be embedded in $O(n,n;\ZZ)$ via
$a\mapsto \begin{pmatrix}a&0\\0&(a^t)^{-1}\end{pmatrix}$ (see
\cite[\S2.4]{GPR}). That means that when $n=1$, $-1\in GL(1,\ZZ)$
should correspond to $(-1, -1)\in O(n,n;\ZZ)$, and the whole
group is generated by this element and by the T-duality
element $\#=\begin{pmatrix}0&1\\1&0\end{pmatrix}$. In fact,
there is an action on $R$ of the larger group $GO(1,1;\ZZ)$, 
generated by $O(1,1;\ZZ)$ and by the matrix
$\begin{pmatrix}-1&0\\0&1\end{pmatrix}$ of determinant $-1$.
This latter element acts by sending $\bc$ to $-\bc$,
$\bch$ to itself, and reversing the sign of the generator of
$\pi_3(R)$. (The latter compensates for the change of sign in the
$k$-invariant $\bc\cup\bch$.) On the level of the universal bundle
space $\bE$, this amounts to a change of sign in the 
cohomology class $\bh\in H^3(\bE)$.

\subsection{The higher-rank case and the T-duality group}
\label{sec:Tdualitygroup}

In the higher rank case where $T$ is a torus of dimension $n$, $n$
arbitrary, we can mimic the construction of \cite{BS} as follows.
(A very similar, but not identical, construction can be found
in \cite{BS2}.)  Let $Z$ be a (nice enough)
space, say a locally compact Hausdorff 
space with the homotopy type of a finite CW complex,
and fix $n \ge 1$.  A \emph{pair} over $Z$ will mean a principal
$T$-bundle $E \to Z$ together with a class $\alpha \in H^3(E,\ZZ)$,
whose restriction to each $n$-torus fiber is $0$.  (The reason for this
extra condition is that, as we showed in Theorem \ref{thm:imageofF},
this condition is necessary and sufficient for getting a compatible $G=\RR^n$
action on $CT(E, \alpha)$.  It also makes the set of pairs over $S^i$
trivial for large $i$, since for $i$ large, all torus bundles over
$S^i$ are trivial, but one has non-trivial $3$-cohomology classes,
coming from the fiber on such bundles.)  
The Bunke-Schick argument from \cite{BS}, copied over essentially
line by line, proves the following. (See also \cite{BS2}, which we
only saw after completing the first draft of this paper.)
\begin{theorem}
\label{thm:classspace}
The set of pairs modulo isomorphism is a 
representable functor, with representing space
\begin{equation}
\label{eq:ClassSpace} 
R = ET \times_T \Maps_0(T, K(\ZZ, 3)), 
\end{equation}
where $ET \to BT = K(\ZZ^n, 2)\cong (\CC\PP^\infty)^n$ 
is the universal $T$-bundle,
and where $\Maps_0$ denotes the set of null-homotopic maps
{\lp}those giving the trivial class in $H^3(T,\ZZ)${\rp}.  
Note that using $\Maps_0$ in place of $\Maps$ makes $R$ 
path-connected.  {\lp}When $n=1$, $\Maps(T, K(\ZZ, 3))$
is the free loop space of $K(\ZZ, 3)$, and is already connected.{\rp}
There is an obvious map
\[
c\co R = ET \times_T (\cdots) \to ET \times_T * = BT 
\]
which corresponds to forgetting the second entry of a pair and just
taking the underlying bundle. This map is a fibration with fiber
$\Maps_0(T, K(\ZZ, 3))$.
\end{theorem}
\begin{proof}
We follow the outline of the argument in \cite{BS}. First we construct
a canonical pair $(\bE, \bh)$ over $R$ by letting $\bE$ be the
pull-back under the map $c$ (defined in the statement of the theorem)
of the universal $T$-bundle $ET\to BT$. Then we define $\bh\co \bE
\to K(\ZZ, 3)$ to be
given by 
\[
\bh(u,\, [v, \gamma]) = \gamma(t),
\]
where $\gamma\in \Maps_0(T, K(\ZZ, 3))$, $u,\,v\in ET$,
$u$ and $v$ live over $c([v, \gamma])$, and $tv=u$. One can check 
that this is independent of the choices of $u$, $v$, and $\gamma$
representing a particular element of $\bE$. Clearly any map $Z\to R$
enables us to pull back the canonical pair $(\bE, \bh)$ to a pair over
$Z$. 

In the other direction, suppose we have a pair $(E, h)$ over
$Z$. Since $E\xrightarrow{p} Z$ is a principal $T$-bundle, we know
that $E\xrightarrow{p}  Z$ is 
pulled back from the universal bundle $ET\to BT$ via a map $f\co Z \to
BT$. We claim we can fill in the diagram
\[
\xymatrix{ &K(\ZZ, 3)& \bE\ar[d] \ar[l]_(.35)\bh \ar[dl]\\
E \ar[d]^p \ar[ru]^h \ar[r]_{\wf}  \ar@{.>}[rru]^{\wp}
& ET \ar[d] & R \ar[ld]^(.45)c\\
Z \ar[r]_(.45)f  \ar@{.>}[rru]^(.4){\varphi}  |\hole& \,BT . &}
\]
as shown to make it commute and to realize $(E, h)$ as the pull-back
of $(\bE, \bh)$. Indeed, we simply define $\varphi(z)=[\wf(e), \gamma]$,
where $e\in p^{-1}(z) \subseteq E$, and where $\gamma\in \Maps(T,
K(\ZZ, 3))$ is defined by $\gamma(t)=h(t\cdot e)$. Since the
definition of a pair includes the requirement that $h$ be
null-homotopic on each torus fiber, $\gamma$ indeed lies in $\Maps_0(T,
K(\ZZ, 3))$. Note that $\varphi(z)$ is independent of the choice of
$e$. We can define $\wp$ by $\wp(e)=\bigl[\wf(e),[\wf(e), \gamma] 
\bigr]$, $e$ as before. The rest of the proof is as in \cite{BS}.
\end{proof}
The next step is to make a detailed analysis of the homotopy
type of $R$.  
\begin{theorem}
\label{thm:homtypeofR}
The space $R$  has only three non-zero homotopy groups, $\pi_1(R)
\cong  \ZZ^k$ {\lp}$k=\binom n2${\rp}, 
$\pi_2(R)\cong \ZZ^{2n}$, and $\pi_3(R)\cong  \ZZ$. 
Moreover, there is a fibration $\bE\to R\to BT$, 
where $\bE$ is a simple space {\lp}one with 
$\pi_1$ acting trivially on all the
homotopy groups{\rp} homotopy
equivalent to $K(\ZZ, 3)\times \bE_0$, with $\bE_0\to R$ inducing an
isomorphism on $\pi_1$, and with the universal cover of
$\bE_0$ homotopy equivalent to $K(\ZZ^n, 2)$.
\end{theorem}
\begin{proof}
The set of pairs over $S^i$ is easy to compute except when
$i=2$ (the only case when the bundle can be non-trivial).  Thus one
finds that
\[
\begin{aligned}
\pi_1(R) & \cong \ker \bigl( H^3(\TT^n \times S^1, \ZZ) \to H^3(\TT^n,
\ZZ) \bigr) 
\cong H^2(\TT^n, \ZZ) \cong \ZZ^k, \quad k = \binom n 2,\\
\pi_3(R) & \cong \ker \bigl( H^3(\TT^n \times S^3, \ZZ) \to H^3(\TT^n,
\ZZ) \bigr) 
\cong H^3(S^3, \ZZ) \cong \ZZ,\\
\pi_i(R) &\cong \ker \bigl( H^3(\TT^n \times S^i, \ZZ) \to H^3(\TT^n,
\ZZ) \bigr) 
=0, \qquad i\ge 4.
\end{aligned}
\]
It follows that $R$ has only three non-zero homotopy groups, $\pi_1(R)
\cong  \ZZ^k$, $\pi_2(R)$, and $\pi_3(R)\cong  \ZZ$. Furthermore,
the proof of Theorem \ref{thm:classspace} showed that we have a 
fibration $\bE \to R \to BT$, with $\bE$ homotopy equivalent
to $\Maps_0(T, K(\ZZ, 3))$, with which we replace it (since
only the homotopy type of $\bE$ matters, anyway).

Next, observe that we can split $\bE$ (up to homotopy) as a product
$\bE_0 \times K(\ZZ, 3)$, where $\bE_0 = \Maps^+_0(T, K(\ZZ, 3))$,
the \emph{based} null-homotopic maps from $T$ to $K(\ZZ, 3)$.
Indeed, we can choose a model $K$ for $K(\ZZ, 3)$ which
is an abelian topological group. (See \cite[Ch.\ V]{May}
and \cite{Gun}.) Then $\bE$, as a space of maps into $K$, has a 
natural topological group structure (coming from pointwise
multiplication in $K$) and thus is simple. Furthermore, we can
construct the desired splitting of $\bE$ by sending $f\co T \to K$ (a
typical element of $\bE$) to the pair $(f(1)^{-1}\cdot f\co T \to K,\,
f(1)\in K)$. (Here we are using the multiplication in $K$, and
$f(1)^{-1}\cdot f$ is based since its value at $1_T$ is $1_K$.)

So to compute
$\pi_2(R)$, we just need to compute $\pi_2(\bE_0)$.
But we have
\[
\pi_j(\Maps^+_0(T, K(\ZZ, 3))) = [\Sigma^j T, K(\ZZ, 3)],\quad j>0,
\]
by the usual identification of $\Maps^+(S^j,\Maps^+(T, W))$ with
$\Maps^+(S^j\wedge T, W)$. 

For simplicity we first consider the case
$n=2$. The suspension of the $2$-torus $T^2 = (S^1\vee S^1)\cup_\psi
e^2$ splits as
\[
\Sigma T \simeq S^2 \vee S^2 \vee S^3,
\]
since the attaching map $\psi$ of the $2$-cell is a commutator in
$\pi_1(S^1\vee S^1)$, and is thus stably trivial. So
\[
\begin{aligned}
&[\Sigma^2T, K(\ZZ, 3)]\\ &= [S^3 \vee S^3 \vee S^4, K(\ZZ, 3)]\\
\phantom{x}&\cong 
[S^3, K(\ZZ, 3)]\oplus [S^3, K(\ZZ, 3)]\oplus [S^4, K(\ZZ, 3)]\\
\phantom{x}&\cong 
\bZ\oplus \bZ \oplus 0 = \bZ^2.
\end{aligned}
\]
Thus $\pi_2(\Maps_0(T, K(\ZZ, 3)))\cong \pi_2(\Maps^+_0(T, K(\ZZ, 3)))
\cong \bZ^2$, and from the exact sequence
\begin{equation}
\label{eq:homotopyseq}
\xymatrix{
0=\pi_3(BT) \ar[r] & \pi_2(\Maps_0(T, K(\ZZ, 3))) \ar[r] & \pi_2(R)
\ar@{->>}[r]^{c_*} & \pi_2(BT)=\ZZ^n,}
\end{equation}
we have $\pi_2(R)\cong \bZ^2\oplus \bZ^2=\bZ^4$.

The higher-rank case works similarly, once we have the appropriate
stable splitting of $T^n$. In fact, if $T$ is a torus of any rank $n$,
then $T$ has the same integral homology as 
\[
\bigvee_{k=1}^n \Biggl(\overbrace{S^k \vee  \cdots \vee S^k}^{\binom n
k}\Biggr). 
\]
\noindent From this one can show that $\Sigma
T$ splits as a wedge of spheres, where the number of $S^{k+1}$
summands is the 
binomial coefficient $\binom n k$. This follows from Whitehead's
Theorem once one can construct a homology equivalence from $\Sigma
T$ to the wedge of spheres; but such a
map can be constructed from the coproduct (over all subsets of
$\{1,\cdots, n\}$) of the maps
\[
\begin{aligned}
\Sigma\Biggl( \overbrace{S^1 \times \cdots\times S^1}^n\Biggr)
&\xrightarrow{\Sigma(\text{proj})} 
\Sigma\Biggl( \overbrace{S^1 \times\cdots \times S^1}^k\Biggr)\\
&\xrightarrow{\Sigma(\text{quotient})} 
\Sigma\Biggl( \overbrace{S^1\wedge  \cdots\wedge S^1}^k\Biggr)
=\Sigma S^k = S^{k+1}.
\end{aligned}
\]
(Note that we need to suspend here first of all to be able to
``add'' maps into the wedge of spheres --- recall $[\Sigma T, Y]$
has a group structure for any $Y$, and secondly
because Whitehead's Theorem only
applies to simply connected spaces.) The same argument as before
then shows that 
\[
\pi_2(\Maps_0(T, K(\ZZ, 3)))\cong \pi_2(\Maps^+_0(T, K(\ZZ, 3)))
\cong \bZ^n,
\]
and from the exact sequence \eqref{eq:homotopyseq},
$\pi_2(R)\cong \bZ^n\oplus \bZ^n$.
\end{proof}

The fundamental group $\pi_1(R)$ presents a problem which shows
up in Theorem \ref{thm:main}; if $H^1(Z,\ZZ)\ne 0$, then
not every pair over $Z$ has a classical T-dual. 
For that reason it is convenient to work
with the universal cover $\wR$, which can be viewed as a classifying
space for pairs over \emph{simply connected} spaces.\footnote{Even in
the simply connected case, we need to be careful --- given $Z$ simply 
connected and a map $Z\to R$, it has a lift to a map $Z\to \wR$,
but the lift is not unique, since we can compose with a covering
transformation. So $\wR$ really classifies pairs together
with a choice of lift.} The space $\wR$
has only two non-zero homotopy groups, $\pi_2$ and $\pi_3$, and so it
is a two-stage Postnikov system just like the Bunke-Schick classifying
space for the case $n=1$. Since every pair over a simply connected
space has a (unique) classical T-dual by Theorem \ref{thm:main}, we
expect T-duality to correspond geometrically to an involution on $\wR$.
Not only this, but we can understand the action of the T-duality group
$O(n,n;\bZ)$ as being the automorphism group of the quadratic form
on $H^2(\wR)$ defined by the $k$-invariant of $\wR$. We formalize this
as follows:

\begin{theorem}
\label{thm:geomTduality}
\hfil\par
\begin{enumerate}
\item
The universal cover $\wR$ of the classifying space $R$ of Theorem
\ref{thm:classspace}
is a two-stage Postnikov system 
\[
K(\ZZ,3) \to \wR \to K(\ZZ^n,2)\times K(\ZZ^n,2),
\]
with $\pi_3(\wR)\cong \bZ$ and
with $\pi_2(\wR)\cong \bZ^n\oplus \bZ^n$. The $k$-invariant of
$\wR$ in $H^4(K(\bZ^n,2)\times K(\bZ^n,2),\bZ)$ can be identified with
$x_1\cup y_1 + \cdots + x_n\cup y_n$, where $x_1,\cdots,x_n$ is a basis
for the $H^2$ of the first copy of $K(\bZ^n,2)$ and
$y_1,\cdots,y_n$ is a basis         
for the $H^2$ of the second copy of $K(\bZ^n,2)$. T-duality is 
implemented by a self-map $\#$ of $\wR$, whose square is homotopic to the
identity, interchanging the two copies of $K(\bZ^n,2)$. {\lp}The involutive 
automorphism of $K(\bZ^n,2)\times K(\bZ^n,2)$ interchanging the two
factors preserves the $k$-invariant and thus extends to a homotopy
involution of $\wR$.{\rp}
The action of the T-duality group $O(n,n;\bZ)$ is by automorphisms of 
$\pi_2(\wR)$ preserving the $k$-invariant.
\item
The classifying space $R$ itself is a simple space {\lp}i.e.,
the fundamental group acts trivially on the higher homotopy groups{\rp},
with $\pi_1(R)\cong H^2(T,\ZZ)\cong \ZZ^k$, $k = \binom n 2$.
\item
Given a space $Z$ and a map $u\co Z\to R$ {\lp}representing, by
Theorem \ref{thm:classspace}, a pair $\bigl(p\co E \to Z, \, \alpha \in 
H^3(E,\ZZ)\bigr)$ with $E$ a principal $T$-bundle over $Z$ and
with $\alpha$ restricting to $0$ on the fibers of $p${\rp}, $u$
has a lifting to a map $Z\to \wR$ if and only if $p_!(\alpha)=0$
in $H^1(Z, H^2(T))$.
\end{enumerate}
\end{theorem} 
\begin{proof}
We have already computed the homotopy groups of $R$. We will first
check (2) and (3), then go back and finish the last part of (1),
which concerns the $k$-invariant of $\wR$. Recall by Theorem
\ref{thm:homtypeofR} that we have a fibration
$\bE\to R\to BT$, where $\bE\to R$ induces
an isomorphism on $\pi_1$ and $\bE$ is simple.
Since $BT$ is simply connected, and the fundamental group
of the fiber $\bE$ must act trivially on the homotopy groups
of the base $BT$, it follows that $R$ is simple.

For (3), observe that the obstruction to lifting $u\co Z\to R$ 
to a map $Z\to \wR$ is the induced map $u_*\co \pi_1(Z) \to
\pi_1(R)\cong H^2(T)$, which can be interpreted as an element
of
\[
\Hom (\pi_1(Z), H^2(T)) \cong \Hom (\pi_1(Z)_{\text{ab}}, H^2(T)) 
\cong H^1(Z, H^2(T)) .
\]
Chasing through the various identifications here shows that this map
is precisely $p_!(\alpha)$, if $u$ corresponds to the pair
$(p\co E \to Z, \, \alpha \in H^3(E))$.

To finish the proof of the theorem, we need to check that the
$k$-invariant of $\wR$ is as described. Clearly $H^2(\wR,\ZZ)$
is free abelian with generators $x_1,\cdots,x_n$ dual to the
generators of $\pi_2(BT)$ and generators $y_1,\cdots,y_n$ dual to the
generators of $\pi_2(\bE)$.  Now given a principal
$\TT$-bundle $p\co
X\to Z$ over a space $Z$ and an element $\delta\in
H^3(X, \ZZ)$, we can make it into a pair $(E_j,\alpha)$ in the sense
of this section, by letting $E_j = X\times 
T^{n-1}\xrightarrow{p\circ \text{pr}_1} Z$, where the
$j$-th copy of $\TT$ in $T^n$ acts by the $\TT$-bundle $X\to Z$,
and the other $n-1$ copies act by translation in the second factor.
This obviously defines a natural transformation from $\TT$-pairs
to $\TT^n$-pairs, and thus a map of classifying spaces 
$\mu_j\co R_1 \to R$,
where $R_1$ is the classifying space of \cite{BS}, which is a
fibration
\[
K(\ZZ, 3) \to R_1 \to K(\ZZ, 2) \times K(\ZZ, 2) 
\]
with $k$-invariant $xy$ ($x$ and $y$ the two generators
of $H^2(R_1)$).
It is clear that $\mu_j$ induces an isomorphism on $\pi_3$
and that it lifts to a map $R_1 \to \wR$ that, on
$H^2(\wR)$, kills the canonical generators
$x_k$ and $y_k$ for $k\ne j$, and pulls back $x_j$ to $x$
and $y_j$ to $y$.  It also must pull back the $k$-invariant
of $\wR$ to the $k$-invariant of $R_1$, which is
$xy$.  This shows that modulo $x_k$ and $y_k$ for
$k\ne j$, the $k$-invariant of $\wR$ is $x_jy_j$.  We're
almost done --- we just need to show there are no cross-terms
involving $x_jx_k$, $y_jy_k$, or $x_jy_k$, with $k\ne j$.
But there can be no terms of the form $y_jy_k$, by the
product splitting of $\bE$ in Theorem \ref{thm:homtypeofR},
and similarly there can no terms of the form $x_jx_k$, since these
would be T-dual to terms of the form $y_jy_k$.  Finally, there
can be no terms of the form $x_jy_k$ with $k\ne j$, since
these would give rise to terms of the form $y_jy_k$
after T-dualizing in the $j$-th circle. Thus the $k$-invariant
must be as described.

Finally, it is clear that the automorphism group $O(n,n;\ZZ)$
of the $k$-invariant acts by homotopy automorphisms on $\wR$.
The element of order $2$ defined by
$\begin{pmatrix} 0 & 1_n\\ 1_n & 0\end{pmatrix}$ pulls
back under each $\mu_j$
to the T-duality transformation for circle bundles, and so
deserves to be called the T-duality transformation in general.
One can check that it matches up with the procedure
we have described earlier, of lifting the $T$-action
on the total space of a pair $(p\co X \to Z, \delta)$
to a $G$-action on $CT(X, \delta)$, then looking
at the pair corresponding to the crossed product.
This procedure, in turn, replaces the original
tori in $X$ by the dual tori in $X^\#$.
Indeed, we know that our topological description matches
the analytic one when $n=1$ (since the T-dual in this
case is uniquely characterized), and thus it must also
agree in this higher-rank case, since we can (say in the
``universal example'' of $\wR$) dualize one circle at a time.
\end{proof}
\begin{remark}
\label{rem:GO}
In fact a slightly larger T-duality group acts on $\wR$,
namely $GO(n,n;\ZZ)$. This is the subgroup of $GL(2n,\ZZ)$
that preserves the quadratic form $x_1y_1+\cdots+x_ny_n$
only up to sign; it is generated by $O(n,n;\ZZ)$ and
by the involution fixing the $y$'s and sending $x_j\mapsto
-x_j$. Of course, since such an involution reverses the
sign of the $k$-invariant of $\wR$, to get it to act
on $\wR$ we also have to reverse the orientation on $K(\ZZ, 3)$.
\end{remark}
\begin{remark}
\label{rem:changeofcoord}
We should emphasize that the T-duality element $\#$ of
$\Aut \wR$ (the homotopy automorphisms of $\wR$)
\emph{depends on a choice of basis} in $H^2(\wR,\ZZ)$.
If we were to choose a different basis $u_1,\cdots,u_n,
v_1,\cdots,v_n$ (such that the $k$-invariant is also
given by $u_1\cup v_1 + \cdots + u_n\cup v_n$),
then clearly interchanging the $u$'s and $v$'s would
give another perfectly valid notion of T-duality.
(This was noticed in \cite[\S4]{BS} in somewhat
different language.) This amounts to saying that
for any $g\in GO(n,n;\ZZ)$,
\[
g\begin{pmatrix} 0 & 1_n\\ 1_n & 0\end{pmatrix}g^{-1}
\]
gives \emph{another} choice of T-duality element.
Thus only the \emph{conjugacy class} of $\#$
under the T-duality group is really canonical.
\end{remark}

Suppose
$(\pi:X\to Z,\delta)$ and $(\pi^\#:X^\#\to Z,\delta^\#)$ are T-dual pairs.
Now we can discuss how, in the case where $Z$ is simply connected
or at least has $H^1(Z,\bZ)=0$, the characteristic classes 
$[\pi]$ and $[\pi^\#]$ of
$\pi$ and $\pi^\#$ are related to the H-fluxes $\delta$ and $\delta^\#$.
For this purpose we fix an identification of the $n$-torus $T$
with $\bT^n$, so that we can think of $[\pi]$ and $[\pi^\#]$ as living
in $H^2(Z,\bZ^n)= \bigl( H^2(Z,\bZ) \bigr)^n$.
The formula that follows is the higher-rank substitute for equation (1)
in \cite[\S4.1]{MR}. It should replace the formula given
in the first version of our preliminary announcement of these results 
in \cite{MR1}, which was not completely correct.
\begin{theorem}
\label{thm:Tdualityformula}
Fix an identification of $T$ with $\bT^n$, let $H^1(Z,\ZZ)=0$,
and let $(\pi:X\to Z,\delta)$ and $(\pi^\#:X^\#\to Z,\delta^\#)$ 
be T-dual pairs. Identify the characteristic class $[\pi]$ of
$\pi$ with an $n$-tuple of classes $[\pi]_j\in H^2(Z,\bZ)$.
For each $j=1,\cdots,n$, let $\pi_j:X \to X_j$ be the $S^1$-bundle
defined by the $j$-th coordinate in $T$, and let $p_j:X_j\to Z$ be the
projection, a principal $\bT^{n-1}$-bundle.  In other words,
$p_j$ is defined by the property that 
\[
[p_j] = \bigl([\pi]_1,\cdots, \widehat{[\pi]_j},\cdots, [\pi]_n\bigr).
\]
Define $X^\#_j$, etc., similarly.  Then 
\[
(p_j)^*\bigl([\pi^\#]_j\bigr) = (\pi_j)_!(\delta),\quad \text{and}\quad
(p^\#_j)^*\bigl([\pi]_j\bigr) = (\pi^\#_j)_!(\delta^\#).
\]
\end{theorem}
\begin{proof}
Since we have shown in Theorem
\ref{thm:classspace} and in Theorem
\ref{thm:geomTduality} that $\pi$ is pulled back from the universal
bundle $\wE\to \wR$, with $\delta$ pulled back from the
canonical class $\bh$ generating $H^3(\wE)\cong \bZ$, it is enough to
prove the  theorem in this case. We begin with some details on the topology
of $\wE$ and $\wR$ which are perhaps of independent interest.
\end{proof} 
\begin{lemma}
\label{lem:cohomcalc}
Let $\wR$ be the classifying space for pairs over simply connected
base spaces, as in Theorem \ref{thm:geomTduality}, and let
$(p\co \wE\to \wR,\bh)$ be the canonical pair over $\wR$.
Then with notation as in Theorem \ref{thm:geomTduality}, 
the cohomology ring $H^*(\wR,\bZ)$ is 
\[
\bZ[x_1,\cdots,x_n,y_1,\cdots,y_n]/(x_1y_1+\cdots +x_ny_n) + \text{torsion},
\]
where the $x_j$'s and $y_j$'s are in degree $2$ and
all the torsion is in degrees $5$ or higher.  In particular,
$H^1(\wR)=H^3(\wR)=0$ and $H^2(\wR)$ and $H^4(\wR)$ are torsion-free.
The characteristic class of $p\co \wE\to \wR$ is
$[p]=(x_1,\cdots,x_n)$, and of the T-dual bundle
is $[p^\#]=(y_1,\cdots,y_n)$. The space $\wE$ is homotopy equivalent
to $K(\bZ,3)\times K(\bZ^n,2)$, so its cohomology ring is
\[            
\bZ[y_1,\cdots,y_n,\iota_3]/(\iota_3^2) + \text{torsion},    
\]     
where the $y_j$'s  are in degree $2$ {\lp}pulled back 
from generators of the same name in $H^2(\wR)${\rp}, $\iota_3$ is 
the canonical generator of $H^3(K(\bZ,3), 3)$ in degree $3$, and
all the torsion {\lp}which can be explicitly described
using the Steenrod algebra \cite[Theorem 6.19]{McC}, since it comes
from the torsion in the cohomology of $K(\bZ, 3)${\rp}, 
is in degrees $5$ or higher.  
\end{lemma}
\begin{proof} From the proof of Theorem \ref{thm:classspace}, $\bE$ is the
homotopy fiber of the map $c\co R\to BT$, and can be identified with 
$\Maps_0(T, K(\ZZ, 3))$, which splits as a product of $K(\bZ,3)$ and
$\Maps_0^+(T, K(\ZZ, 3))$. Thus the universal cover $\wE$ of $\bE$ has
the homotopy type of $
K(\bZ,3)\times K(\bZ^n, 2)$, since $\pi_2\bigl(\Maps_0^+(T,
K(\ZZ, 3))\bigr) \cong \bZ^n$. Now consider the Serre spectral
sequences for the fibrations
\[
\begin{aligned}
K(\bZ,3) \to &\wE \to K(\bZ^n, 2)=(\bC\bP^\infty)^n, \\
K(\bZ,3) \to &\wR \to K(\bZ^{2n}, 2)=(\bC\bP^\infty)^{2n}.
\end{aligned}
\]
These both have the form of Figure \ref{fig:specseq}, with the solid
dots representing the non-zero free abelian groups
(except that for the first of these, $2n$ should be replaced
by $n$ in the label on the horizontal axis). (There is also
higher-rank torsion in $H^*(K(\bZ,3))$, starting with
$\operatorname{Sq}^2 \iota_3$ in dimension $5$ \cite[Theorem
6.19]{McC}.)  The diagonal arrow 
in the picture is $d_4$, and sends $\iota_3$ to the $k$-invariant of
the fibration (see \cite[Ch.\ 6]{McC}), which in the case of $\wE$ is
$0$ since the fibration 
is a product, and in the case of $\wR$ is $x_1y_1+\cdots +x_ny_n$ by
Theorem \ref{thm:geomTduality}.
\begin{figure}[hbt]
\includegraphics[height=1.5in]{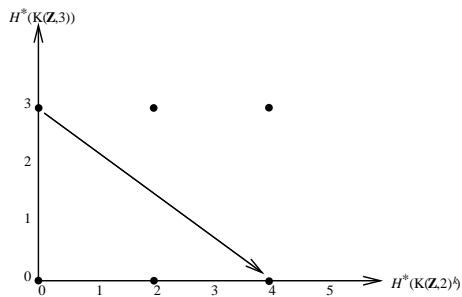}\\
\caption{The Serre spectral sequence}
\label{fig:specseq}
\end{figure}

So modulo torsion starting in dimension $5$, $\wE$ has the same
cohomology ring as $S^3\times(\bC\bP^\infty)^n$, i.e.,
$\bZ[y_1,\cdots,y_n,\iota_3]/(\iota_3^2)$, with $\iota_3$ identified
with the class $\bh$ in Theorem \ref{thm:classspace}. In the case of
$\wR$, since $d_4$ is injective, we see that, again modulo torsion
starting in dimension $5$, the cohomology ring is the same as 
the $E_5$ term in the spectral sequence, 
which is 
\[
\bZ[x_1,\cdots,x_n,y_1,\cdots,y_n]/(x_1y_1+\cdots
+x_ny_n).
\]
\end{proof}
\begin{proof}[Proof of Theorem \ref{thm:Tdualityformula}]
As mentioned above, we just need to compute in the universal
example $\pi\co \wE\to \wR$.  $H^2(\wR)$ was computed in the
Lemma, and $[\pi]=(x_1,\cdots,x_n)$. The H-flux $\delta$ is,
with respect to the notation of Lemma \ref{lem:cohomcalc},
given by $\iota$.  The T-dual bundle
$\pi^\#\co \wE^\#\to \wR$ will have $[\pi^\#]=(y_1,\cdots,y_n)$,
as T-duality interchanges the $x$'s and the $y$'s. The intermediate
bundles $\wE \xrightarrow{\pi_j} \wE_j \xrightarrow{p_j} \wR$
are characterized by
\[
[p_j] = (x_1,\cdots, \widehat{x_j}, \cdots, x_n).
\]
By a calculation identical to that in Lemma \ref{lem:cohomcalc},
we find that 
\[
H^*(\wE_j)\cong \bZ[x_j, y_1, \cdots, y_n]/(x_jy_j)
\] 
(each generator pulled back from one of the same name in $H^2(\wR)$)
modulo torsion in degrees $5$ and higher, and with respect to this
identification, $[\pi_j] = x_j$. Now from the Gysin sequence for
the circle bundle $\pi_j\co\wE\to\wE_j$, we find that
\[
(\pi_j)_!(\iota) = y_j = (p_j)^*([\pi^\#]_j),
\]
as required. The dual formula is now obtained by symmetry.
\end{proof}

\subsection{A splitting of the classifying space when $n=2$}
\label{sec:splitting}
In this subsection, we specialize to the case $n=2$, which was the
subject of our paper \cite{MR}. It is possible that similar
results hold for general $n$, but proving them would be
more complicated.  In any event, our objective here is to show
that, at least for $n=2$, every principal torus bundle with
H-flux is, roughly speaking, decomposed into two pieces: one
with a classical T-dual, and one without.  More precisely,
we have the following rather surprising theorem:
\begin{theorem}
\label{thm:classsplitting}
When $n=2$, if $R$ is the classifying space for pairs
defined in Theorem \ref{thm:classspace} and if $\wR$ is its
universal cover, then $R$ is homotopy equivalent to the
product $\wR\times S^1$.
\end{theorem}
\begin{proof} We have seen that $\pi_1(R)\cong \bZ$, so 
there is a map $\mu\co R\to S^1=K(\ZZ,1)$ which is an isomorphism
on $\pi_1$. This map is necessarily split, since if
$\gamma$ is a generator of $\pi_1(R)$, then by definition,
$\gamma$ corresponds to a map $S^1\to R$ splitting $\mu$.
It remain to construct a map $\beta\co R \to \wR$ which
induces an isomorphism on $\pi_2$ and $\pi_3$; then
$\beta\times \mu$ will be a homotopy equivalence
$R\to \wR\times S^1$. Since $\wR$ is a fibration
\begin{equation}
\label{eq:Post}
K(\ZZ,3) \to \wR \xrightarrow{c_2} K(\ZZ^{2n},2),
\end{equation}
where $c_2$ is a classifying map for $H^2(\wR,\ZZ)$,
we'll begin by constructing a map $R\to K(\ZZ^{2n},2)$
which is an isomorphism on $H^2$, and lift it to a
map to $\wR$ by checking that it's compatible with the
$k$-invariant of \eqref{eq:Post}. To do this, we need
to compute the low-dimensional cohomology of $R$.
We can do this from the spectral sequence
\[
E_2^{p,q}=
H^p(\pi_1(R), H^q(\wR, \ZZ))\Rightarrow H^*(R, \ZZ),
\]
together with the simplicity of $R$ (cf.\ Theorem 
\ref{thm:homtypeofR}), which since the Hurewicz
map $\pi_j(R)\to H_j(R,\ZZ)$ is an isomorphism
for $j=2$ or $3$ by Lemma \ref{lem:cohomcalc}),
guarantees that $\pi_1(R)$ acts trivially on
$H^j(\wR, \ZZ)$, $j\le 3$. Furthermore, from
the same Lemma, the cohomology of $\wR$ is generated
by elements in degrees $2$ and $3$, except perhaps
for some torsion in degree $5$ and up.  So at least
through dimension $4$, $\pi_1(R)$ acts 
trivially on all the cohomology of $\wR$.\footnote{As
a matter of fact, $\pi_1(R)$ acts trivially on \emph{all} of
the cohomology of $\wR$, since the torsion in the cohomology
of $K(\ZZ, 3)$ all comes from the canonical element
$\iota$ via Steenrod operations \cite[Theorem 6.19]{McC}.}
Since $\pi_1(R)\cong \ZZ$,
$E_2^{p,q}=0$ for $p>1$, and thus the spectral sequence
collapses.\footnote{It is this step which would not be
obvious for $n>2$. In the higher-rank case, it is also not
obvious how to construct a map $T^k\to R$ which induces an
isomorphism on $\pi_1$.}  
So, at least modulo torsion in dimensions
$\ge 5$, $R$ has the same cohomology as $\wR\times S^1$.
That means there is a map $R\to  K(\ZZ^{2n},2)$ which
induces an isomorphism on $H^2$, and that this map
lifts to a map $\beta\co R\to \wR$ which is an isomorphism on
$\pi_3$ (since $H^4(R)$ is compatible with the
$k$-invariant of $\wR$). Then
$\beta\times \mu\co R\to \wR\times S^1$ induces
an isomorphism on all homotopy groups, and is
equivariant for the action of the fundamental group
(which is trivial on both sides). Thus it is a
homotopy equivalence by Whitehead's Theorem, since
$R$ has the homotopy type of a CW-complex.    
\end{proof}
\begin{corollary}
\label{cor:conseqofsplit}
For $n=2$, the T-duality group
$GO(2,2;\ZZ)$ acts by homotopy automorphisms, not only on
$\wR$ but also on $R$. For any pair $(p\co X\to Z, \alpha)$
as in Section \ref{sec:Tdualitygroup}, we can attach
two other pairs,  $(p_1\co X_1\to Z, \alpha_1)$ with a 
classical T-dual, and $(p_2\co X_2\to Z, \alpha_2)$
which represents the ``non-classical'' part. 
\end{corollary}
\begin{proof}
This is immediate from the fact that $R$ splits up to
homotopy as $\wR\times S^1$, with the T-duality group
$GO(2,2;\ZZ)$ acting on $\wR$ by homotopy automorphisms.
Given $(p\co X\to Z, \alpha)$, its ``classifying map''
$c\co Z\to R$ is uniquely defined up to homotopy,
by Theorem \ref{thm:classspace}. Then $\mu\circ c:Z\to S^1$
is homotopically non-trivial if and only if $p_!(\alpha)
\ne 0$ in $H^1(Z,\ZZ)$; thus it measures the ``non-classical''
part of the pair.  On the other hand, $\beta\circ c:Z\to 
\wR$ does have a classical T-dual, as described in part
(1) of Theorem \ref{thm:main}.
\end{proof}

\section{The T-duality group in action}
\label{sec:TDinAction}

\subsection{Some classical T-dual examples}
{\em Rank one examples}.
\label{sec:ClassicalAction}

Let $X(p)$ be a circle bundle with $H$-flux
over the 2-dimensional oriented Riemann surface $\Sigma_g$ that has first Chern class
equal to $p$  times the volume form of $\Sigma_g$. For example, when $g=0$, then $\Sigma_g
= S^2$ and $X(p) =  L(1,p)=S^{3}/\ZZ_p$ is the Lens space. 
  The T-duality group in this situation is $GO(1,1; \bbZ)$, and we 
will compute the action of the group, as described in section
\ref{sec:BS}.
 By taking the Cartesian product with a manifold $M$, and pulling back the $H$-flux
to the product, we see that $(M \times X(p), \delta = q)$ 
is T-dual to $(M \times X(q), \delta = p)$, and the element 
$\begin{pmatrix}0 & 1\\ 1& 0\end{pmatrix}$
of $GO(1,1; \bbZ)$ interchanges them. 

The element $\begin{pmatrix}-1 &
0\\ 0& -1\end{pmatrix}$ of the T-duality group 
$GO(1,1; \bbZ)$ lies in the subgroup $GL(1,\ZZ)$, embedded as in section
\ref{sec:BS},
and acts by twisting the $S^1$ action on $M \times X(p)$.
This twisted action makes $M \times X(p)$ into a circle bundle
over $M \times \Sigma_g$ having first Chern class
equal to $-p$ times the volume form of $\Sigma_g$. This bundle is denoted
$M \times X(-p)$, and its total space is diffeomorphic to $M \times X(p)$,
though by an orientation-reversing diffeomorphism.  
Therefore 
the action of $\begin{pmatrix}-1 & 0\\ 0& -1\end{pmatrix}$ on the pair 
$(M \times X(p), \delta=q)$ and $(M \times X(q), \delta=p)$ gives rise to a new T-dual pair
$(M \times X(-p), \delta=-q)$ and $(M \times X(-q), \delta=-p)$. 

The group $GO(1,1; \bbZ)$ is
generated by the two elements of $O(1,1; \bbZ)$ just discussed and by
$\begin{pmatrix}1 & 0\\ 0& -1\end{pmatrix}$, which replaces the
original T-dual pair by the pair consisting of
$(M \times X(p), \delta=-q)$ and $(M \times X(-q), \delta=p)$. 
This refines earlier understood T-duality.
Thus in general there are 8 different (bundle, H-flux) pairs with
equivalent physics, corresponding to $(\pm p, \pm q)$ and
$(\pm q, \pm p)$.
\medskip

\noindent{\em An interesting higher rank example.}

The following interesting test case for our theory was suggested to us
by Professor Edward Witten, who had studied it a few years ago in
joint work with Moore and Diaconescu \cite[end of \S5]{DMW}. 
Namely, let
$X=\bR\bP^7 \times \bR\bP^3$, which is the total space of a principal
$\bT^2$ bundle $\pi$ over 
$Z=\bC\bP^3 \times \bC\bP^1$. The cohomology ring of $Z$ is
$\bZ[c,d]/(c^4,d^2)$, where $c\in H^2(\bC\bP^3)$ and $d\in
H^2(\bC\bP^1)$ are the usual generators.  Note that the characteristic
class $[\pi]\in H^2(Z; \bZ^2)$ that classifies $\pi$ is $(2c, 
2d)$. The cohomology ring of $X$ over the field $\bF$ of two elements
is $\bF[a,b]/(a^8,b^4)$, where $a$ and $b$ are the generators of
$H^1(\bR\bP^7; \bF)$ and of $H^1(\bR\bP^3; \bF)$, respectively. The classes
$a^2$ and $b^2$ are both reductions of integral classes in $H^2$. So
by the K\"unneth Theorem, $H^3(X; \bZ)\cong \bZ\gamma \oplus \bZ/2$, where
the $\bZ/2$ summand is generated by the unique non-zero $2$-torsion class,
the integral Bockstein $\beta(ab)$ of $ab$, and $\gamma$ comes from
$H^3(\bR\bP^3)$. The class $\beta(ab)$ reduces mod 
$2$ to $\operatorname{Sq}^1(ab)=a^2b + ba^2$. Consider the H-flux
$\delta = \beta(ab)$ on $X$ together with the $\bT^2$-bundle $\pi:X\to
Z$. Since $Z$ is simply connected, this data should have a unique
T-dual, and this dual should be classical, i.e., should be given by a
$\bT^2$-bundle $\pi:X^\#\to Z$ with dual H-flux $\delta^\#$, and there
should be an isomorphism on twisted $K$-theory $K^*(X,\delta) \cong
K^*(X^\#,\delta^\#)$.

As we have seen, the T-dual is constructed by taking the crossed
product \\ $CT(X,\delta)\rtimes~ \bR^2$ of the stable continuous-trace
algebra over $X$ with Dixmier-Douady invariant $\delta$ by an
$\bR^2$-action lifting the transitive action of $\bR^2$ on $X$ with
quotient $Z$. If we rewrite the crossed product in two stages as
$\left(CT(X,\delta)\rtimes \bR\right) \rtimes \bR$, we see that we can
compute $X^\#$ in two steps. First we dualize with respect to one
one-dimensional torus bundle, getting $CT(X,\delta)\rtimes\bR\cong
CT(X_1, \delta_1)$, then with respect to the other. We have
a choice of how to factor the action, but let's say we deal first with
the bundle $\bR\bP^7 \to \bC\bP^3$ and then with the bundle $\bR\bP^3
\to \bC\bP^1$. 

Dualizing the first circle bundle gives a diagram
\begin{equation}
\label{eq:firstsq}
\xymatrix{
& \bR\bP^7 \times S^1\times S^3 
\ar[ld]_{\pi_1^*(\pi_2)}\ar[rd]^{\pi_2^*(\pi_1)}\\ 
X=\bR\bP^7 \times \bR\bP^3 \ar[rd]_{\pi_1}& 
& X_1 = \bC\bP^3\times S^1 \times S^3\ar[ld]^{\pi_2}\\
&Z_1 = \bC\bP^3 \times \bR\bP^3&
}
\end{equation}
which we can check and use to compute $\delta_1$
as follows. We already know the identity of the
spaces $X$ and $Z_1$ and of the bundle $\pi_1$. The T-duality (and
Raeburn-Rosenberg) conditions give
\begin{equation} 
\label{eq:firstsqconds}
(\pi_1)_!(\delta) = [\pi_2],\quad
(\pi_2)_!(\delta_1) = [\pi_1]= 2c,
\quad (\pi_1^*(\pi_2))^*(\delta) = (\pi_2^*(\pi_1))^*(\delta_1).
\end{equation} 
In the Gysin sequence for $\pi_1$, the map
$\pi_1^*:H^3(\bC\bP^3 \times \bR\bP^3) \to H^3(X)$ sends the generator
$\gamma$ of $H^3(\bR\bP^3)$ onto itself (if we think of cohomology of the
$\bR\bP^3$ factor as being embedded in the cohomology of either
product by the K\"unneth 
Theorem), so $(\pi_1)_!$ is injective on the torsion in $H^3(X)$ and
sends $\delta$ to a non-zero $2$-torsion class in $H^2(X)$, which must
be $\beta(b)$, the integral Bockstein of $b\in H^1(\bR\bP^3;\bF)$. But
the unique non-trivial principal $\bT$-bundle over $\bR\bP^3$
can be identified with the
short exact sequence of compact Lie groups
\[
\bT \to \text{Spin}^c(3) \twoheadrightarrow SO(3),
\]
where $SO(3)$ is homeomorphic to $\bR\bP^3$ and $\text{Spin}^c(3)$,
the quotient of $\bT\times SU(2)$ by the diagonal copy of $\{\pm 1\}$,
has torsion-free fundamental group, and is thus homeomorphic to
$S^1\times S^3$. That verifies the diagram \eqref{eq:firstsq}.

Now we use the conditions \eqref{eq:firstsqconds} to compute
$\delta_1$. The cohomology ring of $X_1$ 
is $\bZ[c]/(c^4)\otimes\bigwedge_\bZ(\eta,\zeta)$,
where the exterior algebra generators
$\eta$ and $\zeta$ have degrees $1$ and $3$, respectively.
So $H^3(X_1)\cong \bZ c \eta \oplus \bZ\zeta$. 
In the Gysin sequence for $\pi_2$, the map
$\pi_2^*:H^3(\bC\bP^3 \times \bR\bP^3) \to H^3(X_1)$ must be 
injective (since $H^1(\bC\bP^3 \times \bR\bP^3)=0$),
and the image of $(\pi_2)_!$ is the kernel of
cup product with $\beta(b)$, or $2\bZ c \oplus (\bZ/2)\beta(b)$.
In fact, one can see that $(\pi_2)_!(c\eta) = 2c = [\pi_1]$.
So $\delta_1\equiv c\eta$ mod $\ker (\pi_2)_! = 2\zeta \bZ$.

But we also have the condition 
$(\pi_1^*(\pi_2))^*(\delta) = (\pi_2^*(\pi_1))^*(\delta_1)$
from \eqref{eq:firstsqconds}. We have $H^3\bigl(
\bR\bP^7 \times S^1 \times S^3\bigr) \cong \bZ\zeta \oplus (\bZ/2)\beta(a)
\eta$.  To compute $(\pi_1^*(\pi_2))^*(\delta)$, use cohomology with
coefficients in $\bF$: the map $\text{Spin}^c(3) \twoheadrightarrow
SO(3)$ is surjective on $\pi_1$, so the map
$H^*(\bR\bP^3;\bF) \to H^*(S^1\times
S^3, \bF)$ sends the generator $b$ to $\eta$, and
\[
(\pi_2^*(\pi_1))^*(\delta_1)= (\pi_2^*(\pi_1))^*(a^2b + ab^2)
= a^2\eta = \beta(a)\eta,
\]
while $c\eta + 2k \zeta$ pulls back to $\beta(a)\eta + 2k\zeta$.
So this forces $k=0$ and $\delta_1 = c\eta$.  This already has one
surprising consequence, that
\[
K^{*+1}(\bR\bP^7 \times \bR\bP^3, \,\beta(b)) \cong K^*(\bC\bP^3\times
S^1 \times S^3,\, c\eta).
\]
Now the right-hand size can be computed from the twisted
Atiyah-Hirzebruch spectral sequence (see \cite{Ros}) with $E_2$ term
\[
E_2^{p,q} = H^p(\bC\bP^3\times S^1 \times S^3; \bZ),
\qquad q\text{ even (and $0$ for $q$ odd)}
\]
and $d_3=\underline{\phantom{xx}}\cup c\eta$ (since the cohomology is
torsion-free). Now $H^*(\bC\bP^3\times S^1 \times
S^3)=\bZ[c,\eta,\zeta]/(c^4,\eta^2,\zeta^2)$, and
$\ker(\underline{\phantom{xx}}\cup c\eta) = \langle c^3, \eta\rangle$,
$\operatorname{im}(\underline{\phantom{xx}}\cup c\eta) = \langle
c\eta\rangle$. So $E_4=  \langle c^3, \eta\rangle/\langle
c\eta\rangle$ is torsion-free and is generated (as a free abelian
group) by $\eta$, $\eta\zeta$, $c^3$, and $c^3\zeta$.
Thus $K^*(X,\delta) \cong \bZ^2$ in both even and odd degrees.
This seems quite unexpected, since the homology and cohomology of
$\bR\bP^7 \times \bR\bP^3$ has a huge amount of torsion in it.

Now we can go ahead and dualize the other circle bundle. We obtain a
commutative diagram looking like this:
\begin{equation}
\label{eq:secondsq}
\xymatrix{
&& Y
\ar[ld]_{\pi_3^*(\pi_4)}\ar[rd]^{\pi_4^*(\pi_3)}&\\ 
&X_1 = \bC\bP^3\times S^1 \times S^3\ar[ld]^{\pi_2}\ar[rd]_{\pi_3} & &X_2
\ar[ld]^{\pi_4}\\
Z_1 = \bC\bP^3 \times \bR\bP^3\ar[rd]_{\pi_5}&&Z_2\ar[ld]^{\pi_6} \\
& \bC\bP^3 \times \bC\bP^1 &&
}
\end{equation}
Here we should have $\pi_2=\pi_5^*(\pi_6)$ and $\pi_3=\pi_6^*(\pi_5)$.
That means that $[\pi_2]=\beta(b)=\pi_5^*([\pi_6])$, so $[\pi_6]$ must
be an odd multiple of $d\in H^2(\bC\bP^1)$. That in turn means
$Z_2$ must be $\bC\bP^3\times S^3$ or $\bC\bP^3\times L^3$, where
$L^3$ is a $3$-dimensional lens space with finite fundamental group,
but the latter possibility is ruled out by the requirement that 
$S^1\times S^3$ be a principal $S^1$ bundle over $Z_2$.
So $[\pi_6]=d$ and $Z_2 \cong \bC\bP^3\times S^3$. That means
$\pi_3$ is a trivial bundle, so $Y\cong X_2\times S^1$. So the
upper diamond in \eqref{eq:secondsq} becomes
\begin{equation}
\label{eq:thirdsq}
\xymatrix{
& Y = X_2 \times S^1
\ar[ld]_{\pi_3^*(\pi_4)}\ar[rd]^{\text{triv}}&\\ 
X_1 = \bC\bP^3\times S^1 \times S^3\ar[rd]_{\pi_3=\text{triv}} & &X_2
\ar[ld]^{\pi_4}\\
&Z_2= \bC\bP^3\times S^3&
}
\end{equation}
with $[\pi_4] = (\pi_3)_!(\delta_1)= (\pi_3)_!(c\eta) = c$.
Thus $X_2= S^7\times S^3$ with $\pi_4$ the product of the Hopf bundle
$S^7\to \bC\bP^3$ with the identity on $S^3$. Since
the H-flux $\delta_2$ on $X_2= S^7\times S^3$ must pull back to the
pull-back of $c\eta\in H^*(\bC\bP^3\times S^1\times S^3)$ in
$H^*(S^7\times S^1\times S^3)$, $\delta_2=0$, which is also consistent
with our calculation of $K^*(X,\delta)$.

In summary, we have shown that the T-dual of 
$(\pi: \bR\bP^7 \times \bR\bP^3\to Z,\delta)$,
$Z=  \bC\bP^3 \times  \bC\bP^1$ and $\delta = \beta(ab)$,
is the pair $(\pi^\#: S^7\times S^3\to Z,0)$ with trivial 
H-flux and with 
$[\pi]=(2c,2d)\in H^2(Z; \bZ^2)$, $[\pi^\#] = (c, d) \in H^2(Z; \bZ^2)$. 
The element $\begin{pmatrix}1_2 & 0\\ 0& -1_2\end{pmatrix} 
\in GO(2, 2;\bbZ)$ replaces the
original T-dual by the pair consisting of $(S^7 \times S^3, 0)$, where 
$S^3$ and $S^7$ are given the inverse of the standard circle action, 
i.e., we change the signs of the Chern classes of the dual bundle.
An alternate derivation of the T-dual based on physical arguments was given 
in \cite{BEJMS}. 

\section{Possible generalizations}
\label{sec:generalizations}
Instead of just compactifying over torus bundles, one possible
generalization is to compactify
over the classifying spaces $B\Gamma$ of discrete groups $\Gamma$.
(Recall that the rank $n$ torus is $B\bbZ^n$.)
Other examples are Riemann surfaces $\Sigma_g$ of genus $g>0$,
which are classifying spaces of their fundamental groups. 

We will give some evidence for this
here, and begin with a discussion of a special case.
Suppose that $\Sigma = B\Gamma$ is a compact, smooth, connected, manifold, 
where $\Gamma$ is a discrete group.  Also assume for simplicity
that $\Sigma$ has a spin$^c$ structure.  (This is not
essential but will simplify one step in what follows.)
Consider the H-flux 
$\delta =  \tau \times \beta\in H^3(\Sigma \times \TT, \ZZ)$; 
here $\beta \in H^1(\TT, \ZZ)$ is the canonical generator,
and  $ \tau \in H^2(\Sigma,  {\ZZ}) \cong H^2(\Gamma, \ZZ)$ 
determines the central extension
$$
1\to \ZZ \to \Gamma_\tau \to \Gamma \to 1.
$$

\begin{proposition}
Suppose that we are in the situation described above.  
Let $n$ be the dimension of $\Sigma$ {\lp}mod $2${\rp}, so
$n+1$ is the dimension of $B\Gamma_\tau$. Then
there is a commutative diagram,
\begin{equation} \label{eq:Ag}
\begin{CD}
K^\bullet(\Sigma \times \TT, \delta)  @>T_!>> K_{\bullet+n}(C^*(\Gamma_\tau))  \\
       @V{\Ch_\delta}VV          @VV{\Ch} V     \\
H^\bullet  (\Sigma \times \TT,  \delta)    @>T_*>>  HP_{\bullet+n} 
({\mathcal S}(\Gamma_\tau ))
\end{CD}\end{equation}
where $\Ch_\delta$ is the twisted Chern character
and $\Ch$ is the Connes-Chern character. 

If the Baum-Connes 
conjecture holds for the discrete group $\Gamma_\tau$,
then the horizontal arrows are isomorphisms.

\end{proposition} 
\begin{proof}
First observe that if we consider $(\Sigma \times \TT, \delta)$
as a circle bundle with H-flux over $\Sigma$, then its
T-dual is given by the circle bundle $B\Gamma_\tau\to\Sigma$
with first Chern class equal to $\tau \in H^2(\Sigma, \ZZ)$,
with vanishing H-flux on $B\Gamma_\tau$. Thus
T-duality for circle bundles yields the isomorphism
\begin{equation}\label{eqn:1t-dual}
K^\bullet(\Sigma \times \TT, \delta)   \cong K^{\bullet+1}(B\Gamma_\tau).
\end{equation} 
If $\Sigma$ has a 
spin$^c$ structure, then so does $B\Gamma_\tau$ (since it is
the sphere bundle in a complex line bundle over a spin$^c$ manifold).
So by Poincar\'e Duality, $K^{\bullet+1}(B\Gamma_\tau)\cong
K_{\bullet+n}(B\Gamma_\tau)$. Thus the assembly map can be viewed as
a homomorphism 
\begin{equation}\label{eqn:assembly}
K^{\bullet+1}(B\Gamma_\tau) \to K_{\bullet+n}(C^*(\Gamma_\tau)).
\end{equation} 
The composition of the homomorphisms in equations \eqref{eqn:1t-dual}
and \eqref{eqn:assembly}
yields the upper horizontal homomorphism $T_!$;
and the lower homomorphism is similarly defined. 

The validity of the Baum-Connes conjecture for 
$\Gamma_\tau$ is the same as saying that the assembly map given by
equation  \eqref{eqn:assembly}
is an  isomorphism. In this case, it follows that $T_!$ is an isomorphism.
The argument to prove that the
bottom horizontal arrow is an isomorphism is similar, and so is the 
commutativity of the diagram.
\end{proof}

This suggests that the T-dual of $(\Sigma \times \TT, \delta)$ appears to
be the {\Ca} $C^*( \Gamma_\tau)$, which 
can be identified with the algebra of sections of
a continuous field of {\Ca}s over $\TT$, whose fiber
at the point $z\in \TT$ is the twisted group $C^*$-algebra $C^*(\Gamma, z)$.

\section{Open questions}
\label{sec:questions}

We conclude with just a few open questions. First of all, the
``non-classical case'' still remains a bit of a mystery, because of
the lack of symmetry between a torus bundle and its noncommutative
dual. The group $GL(n,\ZZ)$ acts as usual by reparameterizations of
the action, but it's not clear if there is a meaningful $GO(n,n;\ZZ)$
action in this case.

Secondly, we have been unable to determine whether the splitting
result in Theorem \ref{thm:classsplitting} generalizes to the case of
higher rank. An equivalent problem is to determine the Postnikov tower
of $R$ itself, not just that of $\wR$. This is complicated on two
accounts; first of all, one needs to compute the $k$-invariant for the
fibration 
\[
K(\pi_2(R), 2) = K(\ZZ^{2n}, 2) \to R_1 \to K(\pi_1(R), 1) = T^k,
\]
with $k=\binom n 2$,
in $H^3(T^k, \ZZ^{2n})$, and even if this vanishes, one then still
needs to determine if the $k$-invariant of the fibration 
\[
K(\pi_2(R), 3) = K(\ZZ, 3) \to R \to R_1,
\]
in $H^4(R_1,\ZZ)$ lives entirely on the $K(\ZZ^{2n}, 2)$ piece.

\end{document}